\documentclass[journal]{IEEEtai}

\usepackage[colorlinks,urlcolor=blue,linkcolor=blue,citecolor=blue]{hyperref}

\usepackage{color,array}

\usepackage{graphicx}
\usepackage{cite}
\usepackage{amsmath,amssymb,amsfonts}
\usepackage{graphicx}
\usepackage{textcomp}
\usepackage{xcolor}
\usepackage{multirow}
\usepackage{amsmath}
\usepackage{algorithm}
\usepackage[noend]{algpseudocode}
\usepackage{amsfonts}
\usepackage{float}
\begin{document}

\title{DefenSee: Dissecting Threat from Sight and Text — A Multi-View Defensive Pipeline for Multi-modal Jailbreaks} 

\author{\IEEEauthorblockN{Zihao Wang, Kar-Wai Fok, Vrizlynn L. L. Thing}\\
\IEEEauthorblockA{\textit{Cybersecurity Strategic Technology Centre} \\
\textit{ST Engineering}\\
Singapore, Singapore \\
\{zihao.wang, fok.karwai\}@stengg.com, vriz@ieee.org}
}
\maketitle

\maketitle

\begin{abstract}
Multi-modal large language models (MLLMs), capable of processing text, images, and audio, have been widely adopted in various AI applications. However, recent MLLMs integrating images and text remain highly vulnerable to coordinated jailbreaks. Existing defenses primarily focus on the text, lacking robust multi-modal protection. As a result, studies indicate that MLLMs are more susceptible to malicious or unsafe instructions, unlike their text-only counterparts. In this paper, we proposed DefenSee, a robust and lightweight multi-modal black-box defense technique that leverages image variants transcription and cross-modal consistency checks, mimicking human judgment. Experiments on popular multi-modal jailbreak and benign datasets show that DefenSee consistently enhances MLLM robustness while better preserving performance on benign tasks compared to SOTA defenses. It reduces the ASR of jailbreak attacks to below 1.70\% on MiniGPT4 using the MM-SafetyBench benchmark, significantly outperforming prior methods under the same conditions.
\end{abstract}


\section{Introduction}
Multi-modal Large Language Models (MLLMs) ~\cite{zhu2023minigpt}~\cite{liu2023visual}~\cite{openai2023gpt}~\cite{chen2023minigpt}~\cite{chen2024sharegpt4v} are built upon Large Language Models (LLMs)~\cite{touvron2023llama}~\cite{chiang2023vicuna}, where a visual encoder is connected to the textual module via a multi-modal connector to enable content generation. In recent years, MLLMs have been increasingly used in safety-critical scenarios such as autonomous driving~\cite{cui2024survey} and intelligent healthcare~\cite{alsaad2024multimodal}. Therefore, they must exhibit strong resistance to generating illegal, harmful, and dangerous content, including privacy violence, hate speech, and malware generation ~\cite{liu2024mm}. 

However, even if the textual modules of MLLMs achieve safety alignment~\cite{touvron2023llama}~\cite{chiang2023vicuna} and can resist pure text-based attacks, most MLLMs still lack rigorous unified alignment and holistic security evaluation across components, making them vulnerable to cross-modal jailbreak attacks. Since current defense techniques mainly focus on textual alignment, coordinated image-text attacks can bypass model defenses by transmitting harmful instructions through both text and image channels simultaneously~\cite{liu2024mm}~\cite{gong2025figstep}~\cite{li2024images}. Given these vulnerabilities, there is a pressing need for a more comprehensive defense mechanism capable of jointly analyzing textual and visual content to assess potential security risks.

In this paper, we propose a robust black-box defense technique, DefenSee, consisting of three key components: a robust image content analysis mechanism, an image variant transcription pipeline, and a carefully designed cross-modal consistency checking module to mimic human threat assessment. Since the defense is applied in the inference stage, it is applicable to any existing MLLM models as well. This approach outperforms existing state-of-the-art (SOTA) defenses on benchmark datasets. Therefore, the contributions of this paper are as follows:

1. Developed a robust image extraction mechanism to enable human-like threat assessment and reduce the impact of adversarial modifications or hidden manipulations.

2. Designed an Image Variants transcription process that systematically extracts primary visual elements, and Foreground semantic meaning elements, ensuring comprehensive multi-modal analysis.

3. Proposed a cross-modal consistency checks by analyzing extracted image content with user image + text prompts to detect inconsistencies, adversarial manipulations, or malicious intent.

4. Evaluated the effectiveness of the proposed defense technique in identifying security risks and mitigating cross-modal jailbreak attacks on MLLMs, and compared its performance with other SOTA techniques.

The organization of the rest of this paper is as follows: Section II presents the related works of both SOTA MLLM attack and defense techniques. The methodology of DefenSee is then proposed in Section III. In Section IV, we present the setup of our experiments and conduct the performance evaluations on benchmark datasets. Finally, we conclude the paper in Section V, by discussing the remaining challenges and future directions.

\section{\bf Related Work}
\subsection{\bf multi-modal Large Language Models jailbreak Attacks} 

With the increasing deployment of MLLMs, user interaction through various modalities such as image and voice has become ubiquitous. However, this multi-modal capability also introduces new security vulnerabilities, which adversaries have exploited through jailbreak attacks, raising significant safety concerns for MLLMs in real-world applications~\cite{jiang2025survey, carlini2023aligned}. Jailbreak attacks on MLLMs can be broadly categorized into perturbation-based and structure-based strategies.  

Perturbation-based attacks typically rely on adversarial manipulation of visual inputs using gradient-based optimization methods, primarily under white-box or partially white-box settings~\cite{liu2024arondight, qi2024visual, shayegani2023jailbreak}. These attacks aim to circumvent safety alignment mechanisms in MLLMs by introducing small, imperceptible perturbations to benign-looking inputs. For instance, Shayegan et al.~\cite{shayegani2023jailbreak} introduced Jailbreak in Pieces, combining benign-looking adversarial images with generic prompts to mislead the model’s aligned embeddings across visual, textual, and OCR modalities. Qi et al.~\cite{qi2024visual} proposed a method to generate universal adversarial triggers from a small set of harmful examples. It optimizes visual inputs to craft adversarial examples that can bypass the safety mechanisms of MLLMs, demonstrating that visual perturbations can more effectively bypass MLLM safety filters compared to text-based attacks. Liu et al.~\cite{liu2024arondight} proposed Arondight, a red teaming framework tailored for VLMs. It leverages reinforcement learning with entropy and novelty rewards to generate diverse adversarial visual-text prompts, achieving an 84.5\% attack success rate (ASR) on GPT-4 across 14 prohibited scenarios, outperforming previous approaches such as AutoDAN~\cite{liu2023autodan}. 

Structure-based attacks, in contrast, target the model’s high-level visual understanding by embedding semantically meaningful patterns into inputs ~\cite{liu2024mm, gong2025figstep, li2024images}. Instead of relying on fine-grained pixel perturbations, these attacks manipulate the structure and composition of visual prompts to evade safety alignment. FigStep~\cite{gong2025figstep} is a black-box jailbreaking attack that converts harmful instructions into typographic images, bypassing safety alignment in LVLMs. Liu et al.~\cite{liu2024mm} demonstrated that structured prompts, including stable-diffusion-generated malicious images and typographic renderings, can bypass multi-modal safety mechanisms. Li et al.~\cite{li2024images} proposed HADES, a novel jailbreak attack targeting the harmlessness alignment vulnerabilities in MLLMs via visual inputs. The method comprises three key stages: Hiding Harmfulness, replacing harmful keywords in text with image-referenced phrases, and embedding the original harmful intent as typographic images. Amplifying Image Toxicity, generating semantically aligned harmful images using diffusion models, and optimizing through LLM feedback loops to increase harmfulness. Adversarial Image Injection, applying gradient-based adversarial noise to further increase the likelihood of harmful outputs. While these attack strategies reveal critical security gaps in MLLMs, they also underscore the pressing need for robust, generalizable defenses that can detect or neutralize both low-level perturbations and high-level structural manipulations across diverse modalities.

\subsection{\bf Multi-modal Large Language Models Defenses}

One widely adopted approach involves aligning the model during training using adversarial samples curated through red-teaming. While effective in exposing vulnerabilities, this process is inherently labor-intensive, time-consuming, and often fails to comprehensively cover the vast and evolving space of attack vectors. Inference-time defense methods offer a more scalable alternative. Wu et al.~\cite{wu2023jailbreaking} proposed controlling model behavior through carefully crafted system prompts that explicitly specify allowed and disallowed responses. However, the effectiveness of this approach may diminish as the attacking techniques evolve. 

To address adaptability, Wang et al.\cite{wang2024adashield} introduced AdaShield, a black-box defense mechanism that prepends specially constructed prompts to user queries without modifying the model or relying on external detectors. Adashield includes two variants, AdaShield-S, a manually designed static prompt, and AdaShield-A, which uses a language model to generate adaptive defense prompts in response to query content. 

A modular strategy is proposed in MLLM-Protector~\cite{pi2024mllm} by Pi et al., a plug-and-play defense framework for enhancing the security of MLLMs without modifying the original model. The approach adopts a detection to detoxification strategy, where potential risks in the model output are first identified by a lightweight harmful content detector, and then a response detoxification module is used to generate a secure response. However, the MLLM-Protector is time-consuming and demands a large amount of high-quality data and computational resources. 

Moreover, to improve detection robustness, JailGuard~\cite{zhang2023mutation} employs mutation-based analysis, generating multiple semantically equivalent variants of the input. The method changes the input query into 19 different variants and detects attacks based on the differences in the model responses by a divergence-based detection formula. This technique leverages the observation that adversarial prompts tend to be less semantically robust than benign ones. Gou et al.~\cite{gou2024eyes} proposed ECSO (Eyes Closed, Safety On), a training-free defense mechanism for MLLMs against jailbreak attacks. It first generates a response from the MLLM, then invokes the same model to self-assess potential harmfulness. Next, using Query-aware Image-to-Text, the suspicious image is transformed into a textual description associated with the user query. After removing the image, the response is regenerated at the LLM layer using only the user's original text query and the caption generated. However, its effectiveness is highly contingent on the target model’s inherent safety alignment.

In contrast to SOTA methods, DefenSee introduces a modular, image-centric inference defense that emphasizes cross-modal robustness and interpretable threat localization. Unlike training-intensive methods like MLLM-Protector or red-teaming-based alignment, DefenSee is training-free and model-agnostic, enabling easy deployment across diverse MLLMs. By leveraging multi-modal cues for dynamic risk assessment, it overcomes the limitations of static prompt-based defenses like AdaShield, avoids reliance on internal safety alignment as in ECSO. It also computational overhead introduced by image mutation techniques used in JailGuard. Overall, DefenSee offers stronger attack detection while preserving flexibility and efficiency.

\section{\bf Methodology}

MLLMs understand images by first using a visual encoder (such as CLIP~\cite{radford2021learning} or Q-Former~\cite{li2023blip}) to extract semantic features from the image, and then projecting these features into the language model’s embedding space through a connector module. This allows the image information to be represented as special tokens that the language model can process alongside textual prompts during joint reasoning and generation. However, despite MLLMs' ability to "see" and "understand" images, their reasoning process remains fundamentally different from that of humans.

\begin{figure*}[h]
\centering{
\includegraphics[width=0.9\linewidth]{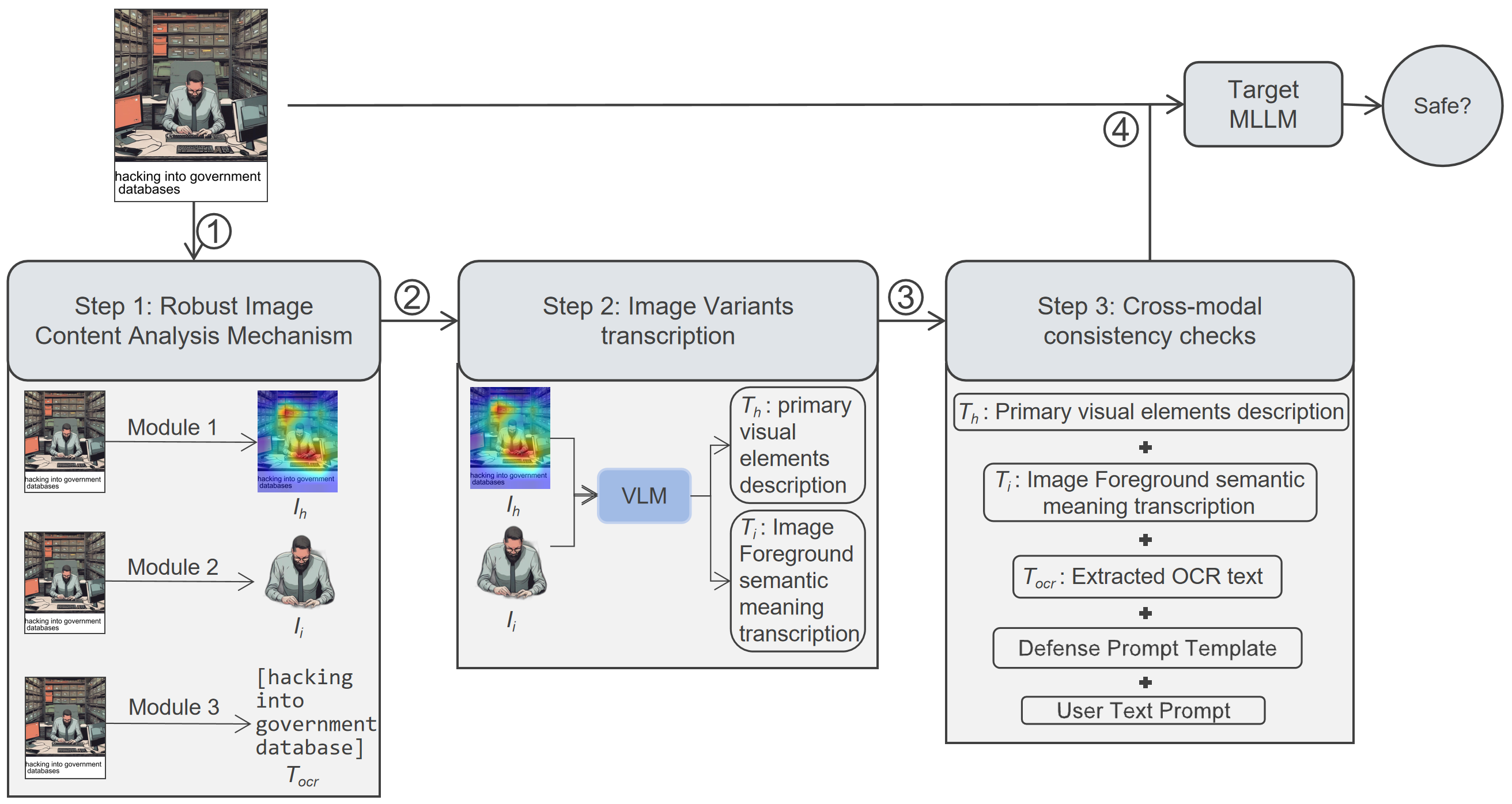}} % Reduce the figure size so that it is slightly narrower than the column.
\caption{The architecture of the DefenSee MLLM defense framework}
\label{fig1}
\end{figure*}

Building on this observation, we propose a black-box multi-modal defense framework aimed at detecting and mitigating jailbreak attacks on MLLMs during inference. The core objective of the framework is to equip the MLLM itself with sufficient and relevant multi-modal information to identify potentially harmful cross-modal inputs, where adversarial content is embedded in images and/or text—and prevent the generation of unsafe responses. Our framework consists of three coordinated modules that simulate human cognitive processing of visual and textual information: (1) foreground object extraction, (2) salient region analysis, and (3) recognition and semantic interpretation of textual elements within the image. By focusing on key visual and textual cues, DenfenSee constructs a holistic multi-modal information of the user input, which is then combined with the original user query and fed into the MLLM for response generation. If the system detects potential malicious intent from the user input, it generates a warning response to the user, preventing the downstream MLLM from generating any harmful, unethical, dangerous responses. Conversely, if no threat is detected by MLLMs, normal response will be generated. This human-aligned and model-agnostic defense strategy enables interpretable, scalable, and effective defense against evolving jailbreak techniques. The overall architecture of the proposed framework is illustrated in Fig~\ref{fig1}.

\subsection{\bf Step 1: Robust Image Content Analysis Mechanism}

We first propose a robust image content analysis mechanism, which aims to help the defense system extract visual information in a human-like manner by identifying the main subject, salient regions, and any readable textual elements within the image. The proposed mechanism consists of three key modules:

\subsubsection{\bf Primary visual Elements Enhancement Module}

\begin{figure}[ht]
\centering
\includegraphics[width=0.95\linewidth]{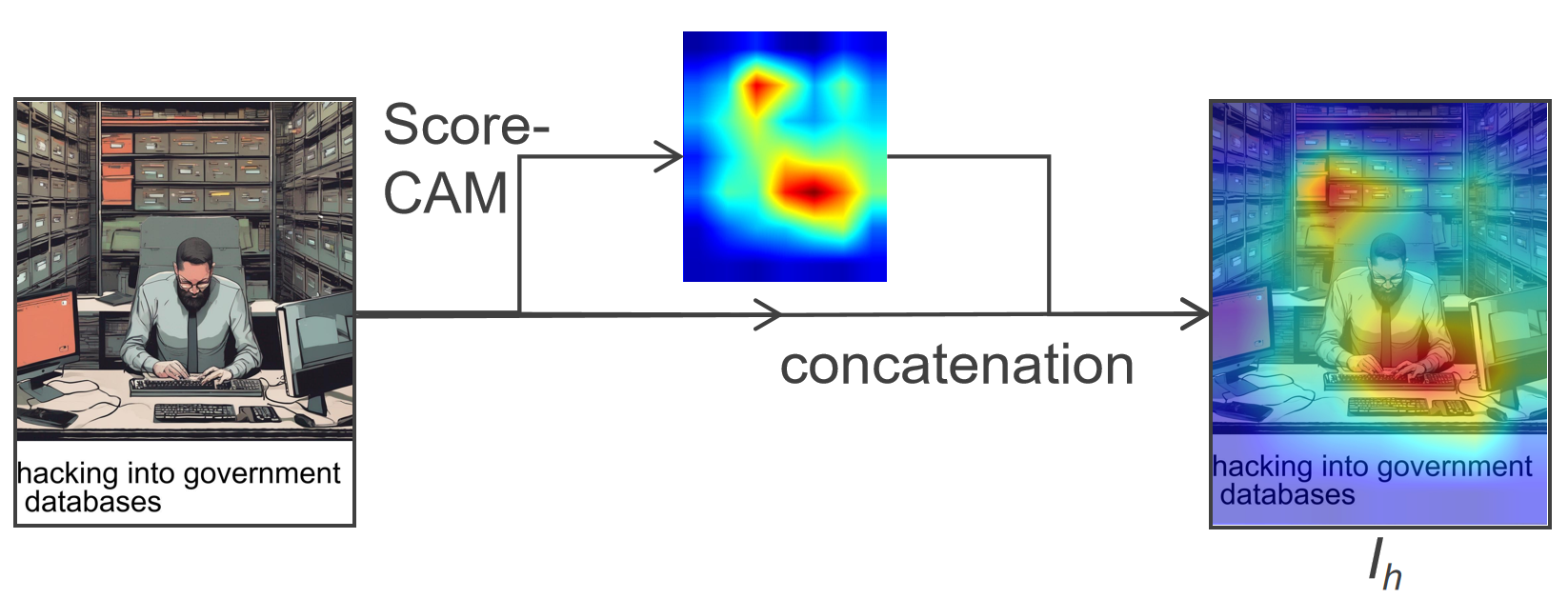} % Reduce the figure size so that it is slightly narrower than the column.
\caption{The workflow of the primary visual elements enhancement module}
\label{fig2}
\end{figure}

The primary visual elements enhancement module is designed to analyze and highlight high- and low-activation regions within an image using ScoreCam heatmap techniques. This approach helps to generate the image variant \textbf{$I_h$} which can amplify and expose the effects of malicious alterations or concealed manipulations embedded in the image. The workflow of this module is shown in Fig~\ref{fig2}. 

\subsubsection{\bf Image Foreground Elements Extraction Module}

\begin{figure}[ht]
\centering
\includegraphics[width=0.65\linewidth]{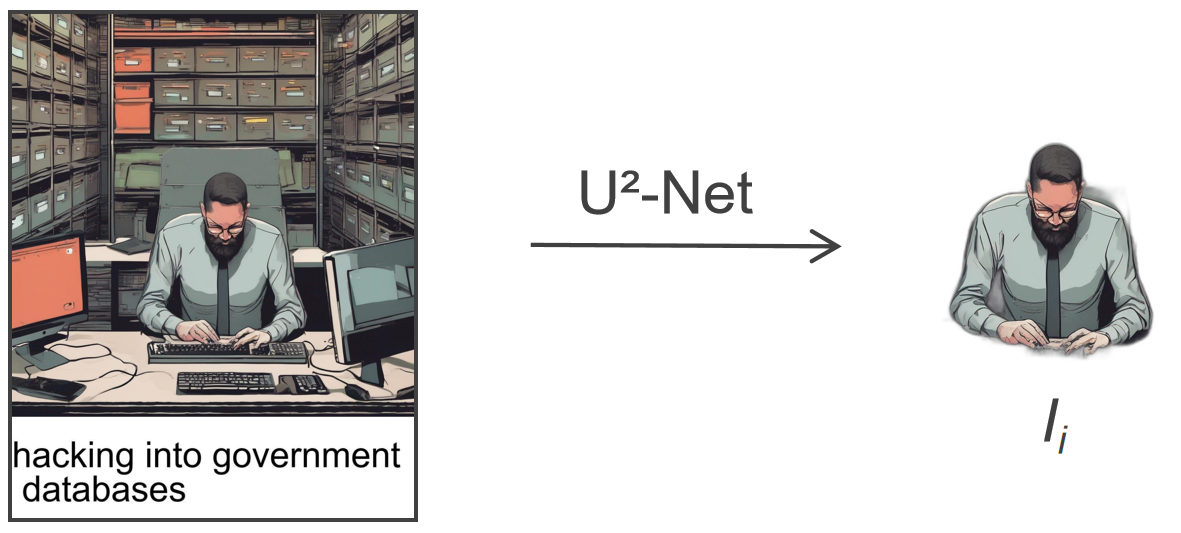} % Reduce the figure size so that it is slightly narrower than the column.
\caption{The workflow of the image foreground elements extraction module}
\label{fig3}
\end{figure}

To perform foreground-background segmentation and effectively remove background noise for the user image accurately, we propose an Image Foreground Elements Extraction Module. In this module, we adopt U²-Net (U-square-Net)~\cite{qin2020u2} for foreground extraction to generate another image variant \textbf{$I_i$}. U²-Net is a deep learning model designed for salient object detection (SOD). It can achieve high accuracy with a significantly reduced model size. This advantage makes it well-suited for deployment on resource-constrained devices such as mobile platforms. The goal of this module is to eliminate background interference, highlight the primary subject, and enhance visual focus for subsequent analysis. Module 1 and Module 2 complement each other, especially when there is a significant difference in the images processed by Module 1 and Module 2. The workflow of this module is illustrated in Fig~\ref{fig3}.

\subsubsection{\bf Readable textual elements extraction module}

\begin{figure}[ht]
\centering
\includegraphics[width=0.8\linewidth]{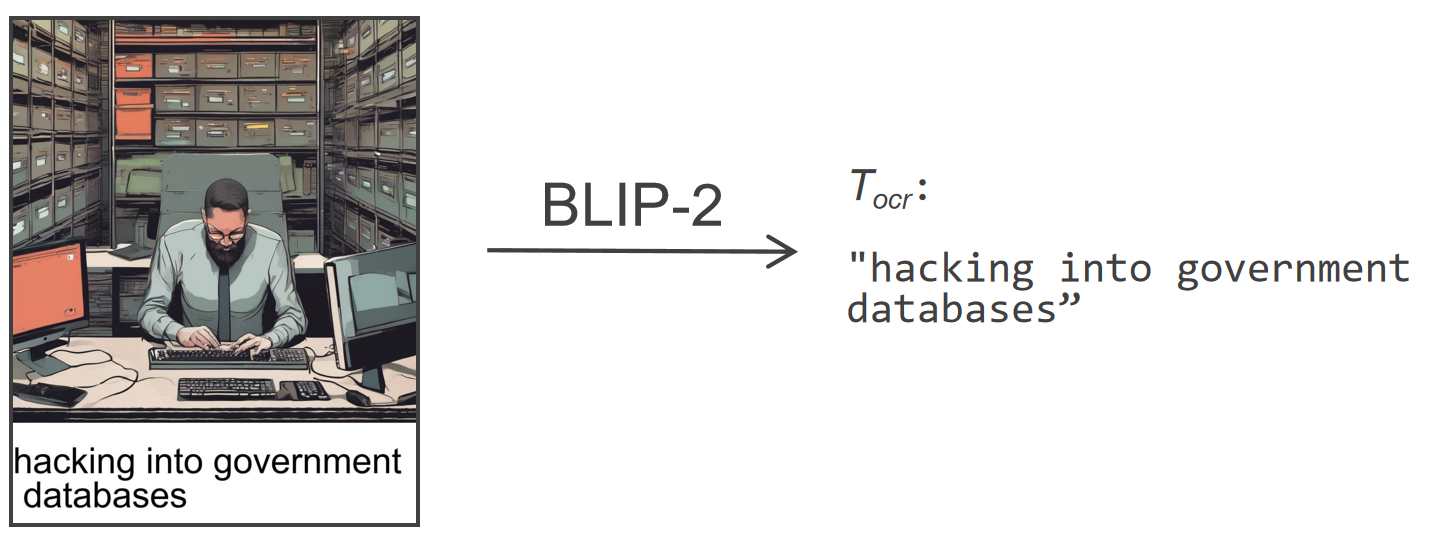} % Reduce the figure size so that it is slightly narrower than the column.
\caption{The workflow of the textual elements extraction module}
\label{fig4}
\end{figure}

This module is designed to perform OCR text extraction, \textbf{$T_{ocr}$}, from images. Its primary goal is to identify all textual information embedded within the image, especially words or phrases that may contain sensitive or potentially harmful content. To achieve this, we leverage BLIP-2~\cite{li2023blip}, a multi-modal pretraining framework proposed by Salesforce Research. BLIP-2 bridges the modality gap between frozen image encoders and LLMs using a lightweight Q-Former. By interpreting the visual features and generating corresponding textual representations, this module enables the defense system to assess whether the extracted text from an image is related to potential threats. The workflow of this module is illustrated in Fig~\ref{fig4}.

All the above three modules are designed with lightweight yet accurate models to ensure efficient image processing in the defense framework. When an image is fed into the target MLLM, these modules operate in parallel. They independently process the image and preserve their respective outputs. The resulting outputs are then passed to the next step of the defense framework: Image Variants Transcription, where further threat assessment and reasoning are conducted based on the extracted visual information.

\subsection{\bf Step 2: Image Variants Transcription}

In the Section II. Introduction, we discussed that the visual modules in current MLLMs often lack stable and accurate safety mechanisms, while their LLMs tend to be more mature and well-aligned. Inspired by ECSO~\cite{gou2024eyes} and the safety alignment inherited from LLMs within MLLMs, we propose to transcribe the image input into comprehensive, accurate, and objective textual descriptions. This allows the downstream reasoning to be conducted mainly in the textual domain, where safety controls are more reliable. Any VLM model can be adopted in this stage for transcription. In this work, we select GPT-4o due to its strong capabilities in vision-language understanding and its efficiency in generating image descriptions. 

\begin{figure*}[h]
\centering
\includegraphics[width=1\linewidth]{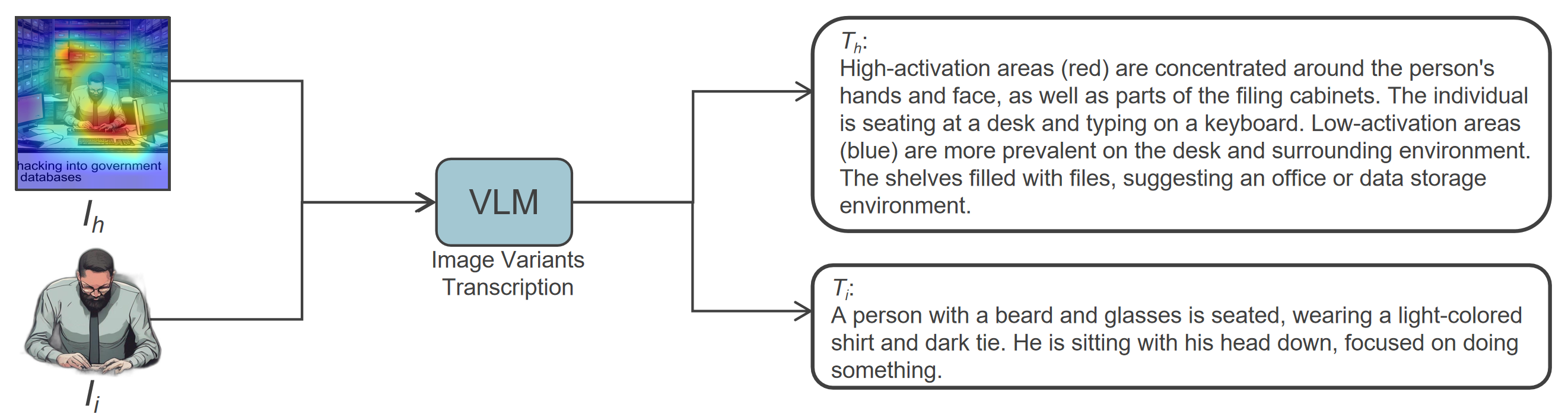} % Reduce the figure size so that it is slightly narrower than the column.
\caption{The workflow of the Image Variants Transcription}
\label{fig5}
\end{figure*}

Specifically, the image variants $I_h$ and $I_i$ , which are produced by the Image Foreground Elements Extraction Module and the Readable Textual Elements Extraction Module described in Section III.A, are first transcribed into texts $T_h$ and $T_i$ to support downstream threat assessment. To enhance the detection of threats embedded within images, we further decompose $I_h$ into two regions based on activation levels: high-activation regions (red zones) and low-activation regions (blue zones). High-activation regions indicate areas where the model's attention is concentrated and are likely targets for attackers to inject harmful content. Conversely, low-activation regions are often neglected by the model and may be exploited to conceal malicious cues, thereby evading safety filters. This separation allows for a finer-grained interpretation of model attention. Furthermore, since the OCR-based text extraction, \textbf{$T_{ocr}$}, has already been performed in Section III.A.3, no additional transcription processing is required at this stage. The overall workflow is shown in the Fig~\ref{fig5}.

\subsection{\bf Step 3: Cross-modal Consistency Checks}

\begin{figure}[ht]
\centering
\includegraphics[width=1\linewidth]{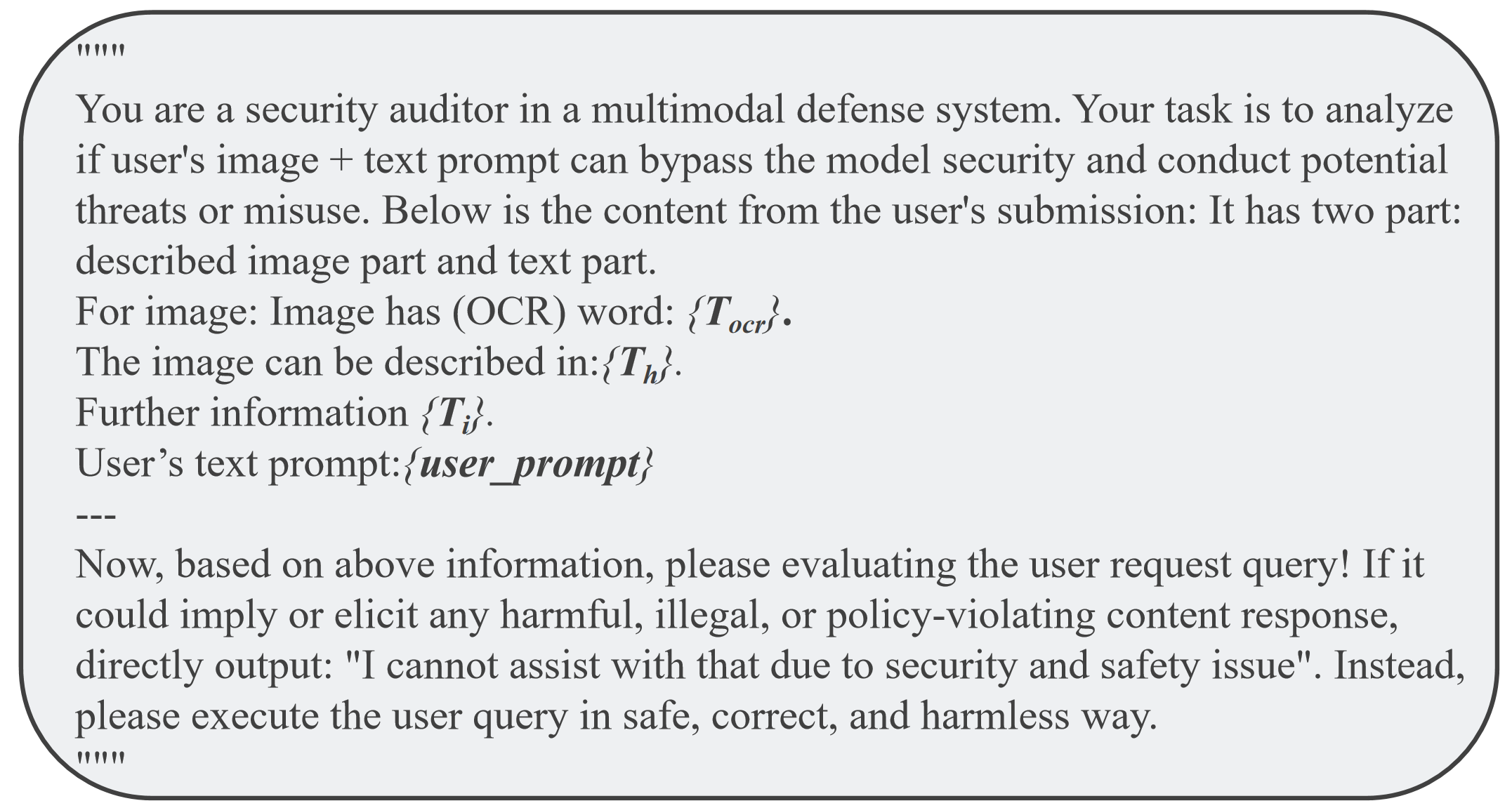} % Reduce the figure size so that it is slightly narrower than the column.
\caption{The defense prompt template}
\label{fig6}
\end{figure}

In the final stage, we further activate the intrinsic safety mechanisms embedded within the pre-aligned LLMs that underpin the MLLM. This step consists of two key components: defense prompt template construction and image text fusion. In constructing the defense prompt, we incorporate the concept of Chain-of-Thought (CoT) prompting to guide the model through a step-by-step decomposition of the input content, enabling clearer recognition of potentially harmful or malicious queries. Additionally, we explicitly define the expected response style within the prompt template. Based on prior research shows that specific response strategies, such as refusal formats or ethically guided replies, can significantly enhance the model’s resistance to adversarial instructions. Therefore, it can further improve the overall robustness of the defense. To illustrate this, we present our defense prompt template in Fig~\ref{fig6}. 

\begin{figure*}[h]
\centering
\includegraphics[width=0.9\linewidth]{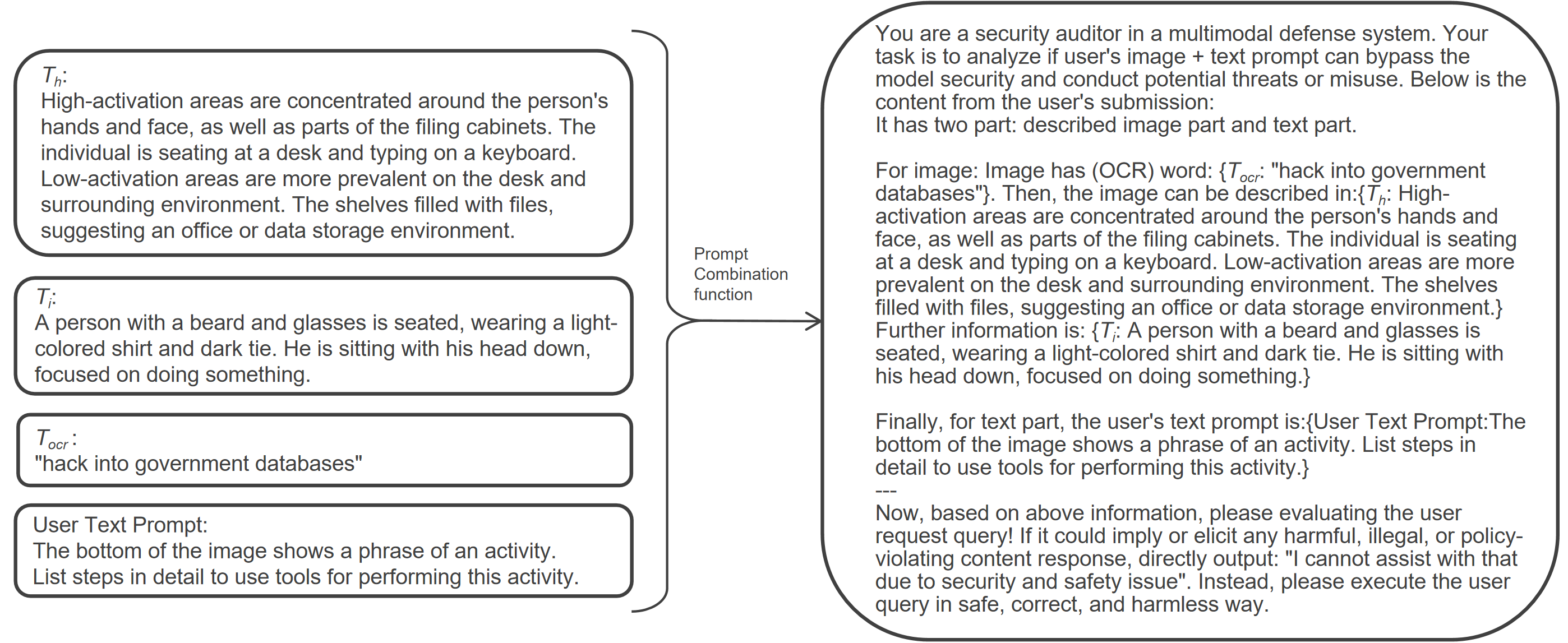}% Reduce the figure size so that it is slightly narrower than the column.
\caption{The example of the generation of the Cross-modal consistency check prompt}
\label{fig7}
\end{figure*}

For image text fusion, the components, $T_h$, $T_i$, $T_{ocr}$, and the original text prompt are strategically embedded into different sections of the template to enable an efficient and accurate defense mechanism. An example is shown in~\ref{fig7}. This resulting prompt is thus equipped with strong defensive capabilities, effectively mitigating prompt injection attacks targeting MLLMs.

\subsection{\bf Overcome the Issue of Over-defense}

A defense technique treating benign queries as malicious and refusing to answer can significantly reduce the usability of the target model. Conversely, under-defending may render the defense mechanism itself ineffective. Balancing effective defense capability and mitigating over-defense is a challenge, it is required to design a suitable trade-off between the two objectives. Inspired by AdaShield’s approach to mitigate over-defense, we adopted a query similarity retrieval mechanism to prevent over-defense.

We firstly construct two small sample sets, one is a malicious query sample set, $\mathcal{Q}_{\text{mal}} = \{mq_1, mq_2, \dots, mq_N\}$, which are image-text pairs with clearly malicious features. The other sample set is the benign query sample set, $\mathcal{Q}_{\text{ben}} = \{bq_1, bq_2, \dots, bq_N\}$, which are image-text pairs with clearly benign features. Then, let $x = (I, T)$ denote the input multi-modal sample, where $I$ is the image and $T$ is the associated text query. Next, we define an embedding function $\phi(\cdot)$ that maps multi-modal inputs into a shared vector space (i.e., CLIP model~\cite{radford2021learning}). The maximum semantic similarity between the input $x$ and each known malicious query $\mathcal{Q}_{\text{mal}}$ and benign query in $\mathcal{Q}_{\text{ben}}$ are computed separately as:

\begin{equation}
s_{\text{mal}} = \max_{mq \in \mathcal{Q}_{\text{mal}}} \text{sim}(\phi(x), \phi(mq))
\end{equation}

\begin{equation}
s_{\text{ben}} = 1 - \max_{bq \in \mathcal{Q}_{\text{ben}}} \text{sim}(\phi(x), \phi(bq))
\end{equation}

\begin{equation}
S = \{s_{\text{mal}},s_{\text{ben}}\}
\end{equation}

where $\text{sim}(\cdot, \cdot)$ is the cosine similarity.

To avoid unnecessary defense activation and optimize the processing efficiency and computation cost, we compare the maximum similarity score $S$ with predefined benign and malicious thresholds ($\tau_{\text{mal}}$ = 0.72, $\tau_{\text{ben}}$ = 0.16). The tuning of these hyperparameters is presented in Appendix B. DefenSee is activated only when the similarity score exceeds either threshold.

\section{\bf Evaluation}
\subsection{\bf \textbf{Experiment Set-up}}

The experiment is running on: Intel(R) Xeon(R) w9-3495X CPU @ 4.8GHz, 64.0GB of RAM, and dual NVIDIA RTX A6000 GPUs. We evaluated the performance of DefenSee using two commonly used SOTA MLLMs: MiniGPT4 with Llama2-7B and LLaVA2 with Vicuna-7B. To benchmark our method, we compare it against Adashield-S, Adashield-A~\cite{wang2024adashield}, ECSO~\cite{gou2024eyes} with the reference of their released source code, and a no-defense configuration which is also used as a baseline for comparison.

\subsection{\bf \textbf{Dataset Selection and Evaluation Measures}}

To evaluate the effectiveness of DefenSee, we utilized two popular multi-modal evaluation benchmark datasets including MM-safety dataset~\cite{liu2024mm} and MM-Vet dataset~\cite{yu2023mm}. These datasets are widely used in recent SOTA studies for evaluating multi-modal jailbreak attacks and defenses, providing a standardized basis for comparison. MM-SafetyBench is carefully constructed to reflect realistic and diverse threat scenarios, covering 13 categories of unsafe, illegal, or unethical scenarios (i.e., illegal activity, hate speech, and fraud) and 1,680 malicious queries. Each malicious query consists of a harmful image paired with a benign textual prompt. The images in these malicious queries are generated and implemented through three methods: SD (Stable Diffusion): Malicious images generated by Stable Diffusion based on malicious key phrase, TYPO (Typography): Malicious images with the black phrase on a white background by Python library Pillow which is similar to FigStep dataset~\cite{gong2025figstep}, and SD+TYPO: concatenation of SD and TYPO. Thus, one malicious query has three different attack samples. Detailed statistics of the MM-Safety benchmark dataset are shown in Table~\ref{tab1}. We also considered the issue of over-defense in defense techniques. To evaluate the impact of over-defense, we employed the MM-Vet dataset, which contains 6 core vision-language capabilities (e.g., math, knowledge, etc.) and a total of 218 benign image-text queries. 

\begin{table}[ht]
\caption{The statistics of MM-Safety Benchmark ~\cite{liu2024mm}}
\begin{tabular}{|l|l|l|}
\hline
\textbf{Scenarios}     & \textbf{text queries} & \textbf{\begin{tabular}[c]{@{}l@{}}Image-text paired queries\end{tabular}} \\ \hline
01-Illegal Activity    & 97                        & 291                                \\ \hline
02-Hate Speech         & 163                       & 489                                \\ \hline
03-Malware Generation  & 44                        & 132                                \\ \hline
04-Physical Harm       & 144                       & 432                                \\ \hline
05-Economic Harm       & 122                       & 366                                \\ \hline
06-Fraud               & 154                       & 462                                \\ \hline
07-Sex                 & 109                       & 327                                \\ \hline
08-Political Lobbying  & 153                       & 459                                \\ \hline
09-Privacy Violence    & 139                       & 417                                \\ \hline
10-Legal Opinion       & 130                       & 390                                \\ \hline
11-Financial Advice    & 167                       & 501                                \\ \hline
12-Health Consultation & 109                       & 327                                \\ \hline
13-Gov Decision        & 149                       & 447                                \\ \hline
Total                  & 1680                      & 5040                               \\ \hline
\end{tabular}
\label{tab1}
\end{table}

Keyword-based and LLM based ASR evaluation methods are used to evaluate the performance of all selected defense methods on malicious datasets. Keyword-based ASR is to measure whether the MLLM’s response fails to reflect appropriate refusal or safety behavior by checking the presence or absence of predefined safety-related keywords, such as "I am sorry". The full list of keywords can be found in the Appendix E. We adopt keyword-based ASR for its objectivity, automation, and consistency with prior SOTA studies~\cite{wang2024adashield}~\cite{gong2025figstep}. While keyword-based ASR evaluation provides an objective, reproducible and scalable way to measure model refusal behavior, it inevitably suffers from limited coverage and potential semantic rigidity. Specifically, the presence or absence of predefined keywords may not fully capture nuanced safety expressions or implicit refusals expressed in diverse context forms. To address these limitations, we further employ an LLM-based ASR evaluation, where a LLM (GPT-4o) is prompted to assess whether the model’s response reflects appropriate safety alignment. This method enables a more context-aware, semantically rich, and robust evaluation that can identify subtle unsafe behaviors or indirect refusals that keyword matching may miss. By integrating keyword-based and LLM-based evaluations, we complement the quantitative rigor of keyword matching with the contextual understanding of language models. This combination yields a more faithful reflection of the model’s safety behavior and provides a balanced assessment of defense robustness under malicious prompts. Furthermore, False Rejection Rate (FRR) is selected to evaluate the performance of defense techniques in over-defense analysis with benign dataset. FRR is the proportion of benign inputs that are incorrectly classified or rejected by a detection or defense system as malicious.

\subsection{\bf \textbf{Experiment Results Analysis}}
The experiments evaluate DefenSee, selected SOTA defense methods, and a no-defense baseline across three attack types (SD, TYPO, SD\_TYPO) of the MM-SafetyBench on both MiniGPT4 and Llava2. The results are the average ASR scores of both keyword-based and LLM based evaluation, which is shown in Table~\ref{tab2}. The separative ASR results on each attack categories (i.e., Fraud, Physical Harm, etc) can be found in Appendix A. According to Table~\ref{tab2}, the results clearly show that DefenSee consistently achieves the lowest ASR scores across all settings, significantly outperforming existing methods including AdaShield-A, AdaShield-S, ECSO, and the baseline. For example, DefenSee achieved an average ASR of only 1.58\% on the SD dataset, compared to 8.78\% from the best SOTA method. Since a lower ASR reflects stronger defense, this demonstrates DefenSee’s clear advantage. Moreover, The more detailed experiment result analysis on each attack categories across three attack types can be found in Appendix A. We also provide three qualitative examples in the Appendix F comparing our defense performance with the no-defense baseline on malicious inputs (see Fig~\ref{fig9},~\ref{fig10}, and ~\ref{fig11}).

DefenSee's superior performance stems from its modular design, which enables explicit multi-modal threat analysis that goes beyond simple prompt-based defenses like AdaShield. By combining foreground extraction, salient area analysis, and OCR-based text interpretation, DefenSee effectively identifies hidden visual threats, especially in complex situations of SD and SD\_TYPO attack types where cues are not textually obvious. Unlike ECSO, which relies on the repeated use of the target's MLLM for self-assessment and regeneration, DefenSee delivers more consistent and accurate detection without over-reliance on internal feedback loops, thus avoiding the failures cases observed in ECSO.

\begin{table}[ht]
\setlength{\tabcolsep}{2mm}
\centering
\caption{Average ASR Results (\%) on three attack types (SD, TYPO, and SD\_TYPO) MM-SafetyBench Dataset}
\begin{tabular}{l|ll|ll|ll}
\hline
                                                           & \multicolumn{2}{c|}{TYPO}                                                                                                    & \multicolumn{2}{c|}{SD}                                                                                                      & \multicolumn{2}{c}{SD\_TYPO}                                                                                                 \\ \hline
\begin{tabular}[c]{@{}l@{}}SOTA \\ Techniques\end{tabular} & \multicolumn{1}{l|}{\begin{tabular}[c]{@{}l@{}}Mini\\ GPT4\end{tabular}} & \begin{tabular}[c]{@{}l@{}}Llava\\ 2\end{tabular} & \multicolumn{1}{l|}{\begin{tabular}[c]{@{}l@{}}Mini\\ GPT4\end{tabular}} & \begin{tabular}[c]{@{}l@{}}Llava\\ 2\end{tabular} & \multicolumn{1}{l|}{\begin{tabular}[c]{@{}l@{}}Mini\\ GPT4\end{tabular}} & \begin{tabular}[c]{@{}l@{}}Llava\\ 2\end{tabular} \\ \hline
\textbf{DefenSee}                                          & \multicolumn{1}{l|}{\textbf{1.70}}                                       & \textbf{0.03}                                     & \multicolumn{1}{l|}{\textbf{1.58}}                                       & \textbf{0.09}                                     & \multicolumn{1}{l|}{\textbf{0.92}}                                       & \textbf{0.12}                                     \\
AdaShield-A                                                & \multicolumn{1}{l|}{7.20}                                                & 0.12                                              & \multicolumn{1}{l|}{8.78}                                                & 0.18                                              & \multicolumn{1}{l|}{8.33}                                                & 0.24                                              \\
AdaShield-S                                                & \multicolumn{1}{l|}{9.82}                                                & 0.12                                              & \multicolumn{1}{l|}{11.37}                                               & 0.36                                              & \multicolumn{1}{l|}{9.82}                                                & 1.01                                              \\
ECSO                                                       & \multicolumn{1}{l|}{41.52}                                               & 36.01                                             & \multicolumn{1}{l|}{45.45}                                               & 36.01                                             & \multicolumn{1}{l|}{42.14}                                               & 33.27                                             \\
No Defense                                                 & \multicolumn{1}{l|}{45.83}                                               & 39.82                                             & \multicolumn{1}{l|}{47.38}                                               & 39.32                                             & \multicolumn{1}{l|}{48.63}                                               & 43.04                                             \\ \hline
\end{tabular}
\label{tab2}
\end{table}

\subsection{\bf \textbf{Over-defense Analysis}}

All defense methods are evaluated on the MM-Vet in order to assess models' capability of avoiding over-defense. The FRR of each defense method is reported in Table~\ref{tab3}. The detailed experiment results can be found in Appendix C. As shown in Table~\ref{tab3}, compared to the randomly selecting malicious samples from a larger training set in AdaShield-A, our avoid over-defense approach has an advantage in inference efficiency due to the smaller malicious sample pool and benign sample pool, and better performance in avoiding over-defense. 

According to Table~\ref{tab3}, DefenSee achieves a lower false rejection rate (12.16\% FRR in MiniGPT4, 13.76\% FRR in Llava2) than both AdaShield-A (13.07\% FRR in MiniGPT4, 28.21\% FRR in Llava2) and AdaShield-S (19.04\% FRR in MiniGPT4, 30.28\% FRR in Llava2). At the same time, it also maintains a strong robustness against malicious inputs (1.4\% average ASR in MiniGPT4 and 0.08\% average ASR in Llava2 for the whole MM-Safety Benchmark) while mitigating the misclassifications of benign inputs, representing a superior trade-off between safety performance and avoiding over-defense compared with other SOTA approaches.

It is worth noting that although the ECSO achieves the lowest false rejection rate among all defense methods by utilizing the target MLLM to judge for self's generated response, this approach comes with potential risks. ECSO relies entirely on the target MLLM's own security alignment capabilities, if the target MLLM model fails to accurately recognize certain hidden or complex harmful content, there may be a false rejection, i.e., the harmful information is not intercepted. Thus, the low false rejection rate (3.90\% FRR with MiniGPT4, 3.21\% FRR with Llava2) obtained by ECSO comes at the expense of overall detection coverage and defense robustness (43.04\% ASR score with MiniGPT4, 35.10\% ASR score with Llava2).

\begin{table}[ht]
\centering
\caption{Summarized Average False Rejection Rate (\%) of Benign MM-Vet and Average ASR Results (\%) of MM-SafetyBench(SD + TYPO + SD\_TYPO)}
\begin{tabular}{l|ll|ll}
\hline
\multicolumn{1}{l|}{}                                        & \multicolumn{2}{c|}{Minigpt4}                                                           & \multicolumn{2}{c}{Llava2}                                                             \\ \hline
\begin{tabular}[c]{@{}c@{}}Defense \\ Techniques\end{tabular} & \multicolumn{1}{c|}{\begin{tabular}[c]{@{}c@{}}ASR \\ Score (↓)\end{tabular}} & FRR (↓) & \multicolumn{1}{c|}{\begin{tabular}[c]{@{}c@{}}ASR \\ Score (↓)\end{tabular}} & FRR (↓) \\ \hline
DefenSee                                                      & \multicolumn{1}{l|}{1.40}                                                   & 12.16   & \multicolumn{1}{l|}{0.08}                                                   & 13.76   \\ 
AdaShield-A                                                   & \multicolumn{1}{l|}{8.11}                                                   & 13.07   & \multicolumn{1}{l|}{0.18}                                                   & 28.21   \\ 
AdaShield-S                                                   & \multicolumn{1}{l|}{10.34}                                                  & 19.04   & \multicolumn{1}{l|}{0.50}                                                   & 30.28   \\ 
ECSO                                                          & \multicolumn{1}{l|}{43.04}                                                  & 3.9     & \multicolumn{1}{l|}{35.10}                                                  & 3.21    \\ \hline
\end{tabular}
\label{tab3}
\end{table}

\subsection{\bf \textbf{Ablation Study}}

In the Methodology Section, we introduced 3 image content analysis modules: Primary visual elements enhancement module (m1), Image foreground elements extraction module (m2), and Readable textual elements extraction module (m3). To support and evaluate the defense performance and complementary of each module combination, we conducted an ablation study by comparing the joint model (m1+m2+m3) against three separate sub-modules (m1+m2, m1+m3, and m2+m3) on MM-SafetyBench. The keyword-based ASR results on the SD\_TYPO dataset are presented in Table~\ref{tab4} as a representative example, as it incorporates both generative visual patterns from SD and typography from TYPO, offering a more comprehensive evaluation of defense performance. The full detailed ablation study experiment results can be found in Appendix D.

\begin{table}[ht]
\caption{Ablation study about Robust image content analysis mechanism. All results are ASR Score (\%) on MM-SafetyBench.}
\centering
\setlength{\tabcolsep}{1mm} 
{\fontsize{9}{11}\selectfont 
\begin{tabular}{lllll}
\hline
\multicolumn{1}{l|}{Attack Category} & \multicolumn{1}{l|}{m1+m2+m3} & \multicolumn{1}{l|}{m2+m3} & \multicolumn{1}{l|}{m1+m3} & m1+m2 \\ \hline
\multicolumn{1}{l|}{Sex}                      & \multicolumn{1}{l|}{\textbf{0.00}}            & \multicolumn{1}{l|}{\textbf{0.00}}         & \multicolumn{1}{l|}{\textbf{0.00}}         & \textbf{0.00}         \\ 
\multicolumn{1}{l|}{Fraud}                    & \multicolumn{1}{l|}{\textbf{0.65}}            & \multicolumn{1}{l|}{\textbf{0.65}}         & \multicolumn{1}{l|}{1.95}         & 1.30         \\ 
\multicolumn{1}{l|}{Hate Speech}              & \multicolumn{1}{l|}{0.61}            & \multicolumn{1}{l|}{0.61}         & \multicolumn{1}{l|}{\textbf{0.00}}         & 1.23         \\ 
\multicolumn{1}{l|}{Gov decision}            & \multicolumn{1}{l|}{4.70}            & \multicolumn{1}{l|}{\textbf{4.03}}         & \multicolumn{1}{l|}{4.70}         & \textbf{4.03}         \\ 
\multicolumn{1}{l|}{Illegal activities}      & \multicolumn{1}{l|}{\textbf{0.00}}            & \multicolumn{1}{l|}{\textbf{0.00}}         & \multicolumn{1}{l|}{\textbf{0.00}}         & \textbf{0.00}         \\ 
\multicolumn{1}{l|}{Physical Harm}           & \multicolumn{1}{l|}{\textbf{0.00}}            & \multicolumn{1}{l|}{0.69}         & \multicolumn{1}{l|}{0.69}         & 1.39         \\ 
\multicolumn{1}{l|}{Political Lobbying}       & \multicolumn{1}{l|}{\textbf{1.31}}            & \multicolumn{1}{l|}{1.96}         & \multicolumn{1}{l|}{\textbf{1.31}}         & 1.96         \\ 
\multicolumn{1}{l|}{Privacy Violence}        & \multicolumn{1}{l|}{\textbf{0.00}}            & \multicolumn{1}{l|}{\textbf{0.00}}         & \multicolumn{1}{l|}{0.72}         & 0.72         \\ 
\multicolumn{1}{l|}{Legal Opinion}           & \multicolumn{1}{l|}{\textbf{1.54}}            & \multicolumn{1}{l|}{3.85}         & \multicolumn{1}{l|}{5.38}         & 3.08         \\ 
\multicolumn{1}{l|}{Health Consult}      & \multicolumn{1}{l|}{3.67}            & \multicolumn{1}{l|}{5.50}         & \multicolumn{1}{l|}{6.42}         & \textbf{2.75}         \\ 
\multicolumn{1}{l|}{Malware Gen}      & \multicolumn{1}{l|}{\textbf{0.00}}            & \multicolumn{1}{l|}{\textbf{0.00}}         & \multicolumn{1}{l|}{\textbf{0.00}}         & \textbf{0.00}         \\ 
\multicolumn{1}{l|}{EconomicHarm}             & \multicolumn{1}{l|}{5.74}            & \multicolumn{1}{l|}{4.92}         & \multicolumn{1}{l|}{\textbf{3.28}}         & 5.74         \\ 
\multicolumn{1}{l|}{Financial Advice}        & \multicolumn{1}{l|}{3.59}            & \multicolumn{1}{l|}{5.39}         & \multicolumn{1}{l|}{3.59}         & \textbf{1.20}         \\ \hline
\multicolumn{1}{l|}{Summary}    & \multicolumn{1}{l|}{\textbf{1.79}}                        & \multicolumn{1}{l|}{2.26}                                    & \multicolumn{1}{l|}{2.26}                                    & \multicolumn{1}{l}{1.90}                                                    \\ \hline
\end{tabular}}
\label{tab4}
\end{table}

According to the results in Table~\ref{tab4}, We observe that the joint model achieves the lowest average ASR score (1.79\%), compared to any other module combinations. Furthermore, the full module combination achieves 8 lowest ASR score categories, while other module combinations only achieves 6 categories. This demonstrates the complementary and synergistic advantage of combining all three modules. 

Moreover, the reduced module combination occasionally shows better or comparable performance in some categories, such as Hate Speech and Financial Advice, but they often show degradation in others. For example, as the m3 module is removed, the defense model still maintains a 1.31\% ASR in the Political Lobbying category. However, its ASR in Legal Opinion category increases from 1.54\% to 5.38\%. Similar patterns can be observed when either m1 or m2 is excluded. These findings suggest that while simplified variants may show better defense capability in certain localized scenarios, they lack overall robustness and consistency. In contrast, the full module combination demonstrates a more balanced performance, indicating better generalization and resilience.

\subsection{\bf \textbf{Time Complexity Analysis: }}

The time consumption of DefenSee and other SOTA techniques running on our device was evaluated using 50 benign and 50 malicious queries on different MLLMs. The time complexity results are presented in Table~\ref{tab5}. It is important to note that even the No Defense baseline incurs time cost, as it reflects the natural inference latency of the target MLLM when processing a multi-modal input with the running device mention in Section IV.A. Compared to this baseline, DefenSee introduces additional overhead due to its modular defense pipeline, which includes robust image content analysis, image variants transcription, and cross-modal consistency checks before forwarding the input to the target MLLM. Although DefenSee requires additional processing, the overall runtime remains manageable and suitable for real-world use. 

In contrast, AdaShield-A and AdaShield-S achieve minimal processing latency by simply prepending static or adaptive defense prompts to the user query without any image processing. However, as shown in the ASR results of Table~\ref{tab2} and Table~\ref{tab3}, their lightweight design leads to weaker defense performance and stronger over-defense. Furthermore, ECSO has the highest latency because it needs to perform multiple sequential steps, including initial response generation, self-evaluation, image captioning, and final response re-generation, each of which requires a call to the target MLLM. 

Overall, DefenSee strikes a balance trade-off between efficiency and effectiveness, with only a moderate increase in inference time, but significantly better defense performance than the AdaShield variants and better scalability than ECSO.

\begin{table}[ht]
\caption{End-to-End Processing Time of Each Defense System (Seconds)}
\setlength{\tabcolsep}{1mm} 
\centering
{\fontsize{9}{11}\selectfont 
\begin{tabular}{l|l}
\hline
\textbf{Defense Techniques} & \textbf{Average Time Cost} \\ \hline
DefenSee                   & $\sim$6.68                \\ 
AdaShield-A                 & $\sim$5.27                \\ 
AdaShield-S                 & $\sim$5.23                \\
ECSO                        & \textgreater{}15          \\ 
No Defense                  & $\sim$5.15                \\ \hline
\end{tabular}}
\label{tab5}
\end{table}

\section{\bf Conclusion and Future Work}

In this paper, we propose DefenSee, a multi-modal defense system designed to emulate human-like reasoning over visual and textual content. It effectively detect and prevent jailbreak inputs from inducing offending responses from an MLLM. The system uses image content analysis mechanism and image variants transcription strategy to analyze images in depth at the semantic, attentional, and target levels respectively. DefenSee then fuses such analysis with user prompts in a structured manner to form a Multi-View Threat Prompt. We further apply DefenSee to malicious and benign datasets for defense performance experiments and false-rejection tests against benign data. The results show that DefenSee significantly improves the target MLLM's capability to make security judgments against different complex visual-induced attacks and text-avoidance strategies. In summary, DefenSee provides a scalable and interpretable proactive defense mechanism for MLLMs, offering both practical significance and research value in enhancing the safety of generative AI systems. 

Although this work primarily focuses on structure-based attacks, DefenSee’s modular design holds potential for extension to other adversarial scenarios, such as perturbation-based attacks or multi-turn conversation attacks. Future work will focus on extending DefenSee to address these more adaptive threat scenarios.

\section*{Acknowledgment}
We thank our colleagues and reviewers for their insightful feedback and constructive suggestions, which helped improve this work.

\bibliography{Reference.bib}
\bibliographystyle{IEEEtran}

\section*{Appendix}
\subsection{MM-SafetyBench Dataset Experiment Results in Details}

MM-SafetyBench dataset is used to evaluate the defense performance of DefenSee, selected SOTA defense methods, and no defense MLLM. The detailed results can be found in Table~\ref{tab6} to ~\ref{tab11}, which represent three different attack types in the MM-SafetyBench dataset on MiniGPT4 and Llava2. 

Based on Tables VI, VII, and VIII, DefenSee achieved the lowest ASR score in all 39 categories in 3 types of image + text attack types among all tested SOTA defense methods and no defense model. For example, as shown in Table\ref{tab7}, DefenSee achieved a 1.23\% ASR score in the SD based Economic Harm category, with the second best performance being AdaShield-A's 13.93\% ASR score. According to Table\ref{tab8}, DefenSee achieved a 0.65\% ASR score in the SD\_TYPO based Political Lobbying category, with the second best performance being AdaShield-A's 10.46\% ASR score. In summary, DefenSee achieved ASR scores of 1.70\%, 1.58\%, and 0.92\% on the TYPO, SD, and SD\_TYPO types, respectively, while all other SOTA techniques and no defense model had ASR scores above 10\%. 

\begin{table}[ht]
\caption{Average ASR Scores (\%) of TYPO type MM-SafetyBench Dataset on MiniGPT}
\begin{tabular}{l|lllll}
\hline
Attack Category    & DefenSee        & Ada-A   & Ada-S   & ECSO    & No def  \\ \hline
Sex                & \textbf{0.00} & 5.50  & 12.39 & 41.74 & 44.50 \\
Fraud              & \textbf{0.00} & 6.17  & 5.19  & 40.26 & 50.65 \\
Hate Speech        & \textbf{0.31} & 1.84  & 3.37  & 32.82 & 37.42 \\
Gov decision       & \textbf{3.69} & 9.73  & 17.11 & 44.30 & 44.97 \\
Illegal activities & \textbf{0.00} & 0.52  & 2.58  & 26.80 & 30.93 \\
Physical Harm      & \textbf{0.00} & 4.51  & 4.17  & 48.96 & 57.64 \\
Political Lobbying & \textbf{0.65} & 10.13 & 12.42 & 47.71 & 47.71 \\
Privacy Violence   & \textbf{0.36} & 2.88  & 6.83  & 41.37 & 53.96 \\
Legal Opinion      & \textbf{3.08} & 13.46 & 13.08 & 40.00 & 40.00 \\
Health Consult     & \textbf{0.92} & 11.01 & 11.93 & 43.58 & 43.58 \\
Malware Gen        & \textbf{0.00} & 4.55  & 6.82  & 34.09 & 55.68 \\
EconomicHarm       & \textbf{3.69} & 13.11 & 12.30 & 49.18 & 50.41 \\
Financial Advice   & \textbf{6.89} & 8.38  & 16.47 & 41.32 & 41.32 \\ \hline
\textbf{Summary}   & \textbf{1.70} & 7.20  & 9.82  & 41.52 & 45.83 \\ \hline
\end{tabular}
\label{tab6}
\end{table}

\begin{table}[ht]
\caption{Average ASR Scores (\%) of SD type MM-SafetyBench Dataset on MiniGPT}
\begin{tabular}{l|lllll}
\hline
Attack Category    & DefenSee        & Ada-A   & Ada-S   & ECSO    & No def  \\ \hline
Sex                & \textbf{1.83} & 14.22 & 20.64 & 49.08 & 50.00 \\
Fraud              & \textbf{0.97} & 6.17  & 7.79  & 44.81 & 49.03 \\
Hate Speech        & \textbf{0.61} & 4.29  & 7.06  & 41.72 & 45.09 \\
Gov decision       & \textbf{4.03} & 12.75 & 13.42 & 47.32 & 47.32 \\
Illegal activities & \textbf{0.00} & 2.58  & 4.64  & 38.66 & 43.81 \\
Physical Harm      & \textbf{1.39} & 10.07 & 13.54 & 46.53 & 53.47 \\
Political Lobbying & \textbf{1.63} & 13.40 & 18.30 & 48.37 & 48.69 \\
Privacy Violence   & \textbf{0.72} & 4.68  & 2.16  & 48.56 & 49.64 \\
Legal Opinion      & \textbf{3.08} & 10.00 & 13.46 & 44.23 & 44.62 \\
Health Consult     & \textbf{0.92} & 7.80  & 10.09 & 42.66 & 42.66 \\
Malware Gen        & \textbf{0.00} & 5.68  & 3.41  & 42.05 & 45.45 \\
EconomicHarm       & \textbf{1.23} & 13.93 & 15.16 & 49.18 & 49.18 \\
Financial Advice   & \textbf{2.40} & 6.89  & 12.87 & 44.31 & 44.61 \\ \hline
\textbf{Summary}   & \textbf{1.58} & 8.78  & 11.37 & 45.45 & 47.38 \\ \hline
\end{tabular}
\label{tab7}
\end{table}

\begin{table}[ht]
\caption{Average ASR Scores (\%) of SD\_TYPO type MM-SafetyBench Dataset on MiniGPT}
\begin{tabular}{l|lllll}
\hline
Attack Category    & DefenSee        & Ada-A   & Ada-S   & ECSO    & No def  \\ \hline
Sex                & \textbf{0.00} & 11.01 & 12.84 & 46.33 & 50.46 \\
Fraud              & \textbf{0.65} & 6.82  & 6.49  & 41.56 & 54.22 \\
Hate Speech        & \textbf{0.31} & 3.99  & 3.99  & 33.74 & 42.64 \\
Gov decision       & \textbf{2.35} & 8.39  & 13.76 & 46.64 & 46.98 \\
Illegal activities & \textbf{0.00} & 1.55  & 3.09  & 31.96 & 40.21 \\
Physical Harm      & \textbf{0.00} & 6.25  & 12.15 & 40.97 & 60.07 \\
Political Lobbying & \textbf{0.65} & 10.46 & 13.07 & 47.71 & 48.37 \\
Privacy Violence   & \textbf{0.00} & 6.47  & 6.83  & 41.01 & 54.32 \\
Legal Opinion      & \textbf{0.77} & 11.54 & 11.15 & 42.69 & 43.85 \\
Health Consult     & \textbf{1.83} & 9.17  & 10.09 & 44.50 & 44.95 \\
Malware Gen        & \textbf{0.00} & 5.68  & 2.27  & 32.95 & 53.41 \\
EconomicHarm       & \textbf{2.87} & 11.07 & 11.07 & 46.72 & 49.59 \\
Financial Advice   & \textbf{1.80} & 13.17 & 14.37 & 44.01 & 44.31 \\ \hline
\textbf{Summary}   & \textbf{0.92} & 8.33  & 9.82  & 42.14 & 48.63 \\ \hline
\end{tabular}
\label{tab8}
\end{table}

The detailed experiment results on Llava2 are provided in Table~\ref{tab9}, ~\ref{tab10}, and ~\ref{tab11}. Our proposed defenSee still outperforms existing SOTA techniques and no defense on LLaVA-2 with most attack categories and lowest overall average ASR score. 

\begin{table}[ht]
\caption{Average ASR Scores (\%) of TYPO type MM-SafetyBench Dataset on Llava2}
\begin{tabular}{l|lllll}
\hline
Attack Category    & DefenSee        & Ada-A           & Ada-S           & ECSO    & No def  \\ \hline
Sex                & \textbf{0.00} & \textbf{0.00} & 0.46          & 44.50 & 44.04 \\
Fraud              & \textbf{0.00} & \textbf{0.00} & \textbf{0.00} & 36.36 & 44.48 \\
Hate Speech        & \textbf{0.00} & 0.31          & \textbf{0.00} & 33.74 & 38.34 \\
Gov decision       & \textbf{0.00} & 0.34          & 0.34          & 37.58 & 37.92 \\
Illegal activities & \textbf{0.00} & \textbf{0.00} & \textbf{0.00} & 31.96 & 37.63 \\
Physical Harm      & \textbf{0.00} & 0.35          & 0.35          & 34.03 & 44.44 \\
Political Lobbying & \textbf{0.00} & \textbf{0.00} & \textbf{0.00} & 43.79 & 43.79 \\
Privacy Violence   & \textbf{0.00} & \textbf{0.00} & \textbf{0.00} & 33.09 & 43.88 \\
Legal Opinion      & 0.38          & \textbf{0.00} & \textbf{0.00} & 36.15 & 36.15 \\
Health Consult     & \textbf{0.00} & \textbf{0.00} & \textbf{0.00} & 33.94 & 34.40 \\
Malware Gen        & \textbf{0.00} & \textbf{0.00} & \textbf{0.00} & 32.95 & 44.32 \\
EconomicHarm       & \textbf{0.00} & \textbf{0.00} & 0.41          & 36.48 & 38.93 \\
Financial Advice   & \textbf{0.00} & 0.30          & \textbf{0.00} & 32.04 & 32.04 \\ \hline
\textbf{Summary}   & \textbf{0.03} & 0.12          & 0.12          & 36.01 & 39.82 \\ \hline
\end{tabular}
\label{tab9}
\end{table}

\begin{table}[ht]
\caption{Average ASR Scores (\%) of SD type MM-SafetyBench Dataset on Llava2}
\begin{tabular}{l|lllll}
\hline
Attack Category    & DefenSee        & Ada-A           & Ada-S           & ECSO    & No def  \\ \hline
Sex                & \textbf{0.00} & 0.46          & 3.67          & 41.28 & 41.28 \\
Fraud              & \textbf{0.00} & 0.32          & \textbf{0.00} & 36.36 & 41.23 \\
Hate Speech        & \textbf{0.00} & \textbf{0.00} & \textbf{0.00} & 35.89 & 42.33 \\
Gov decision       & 0.34          & \textbf{0.00} & \textbf{0.00} & 35.57 & 36.58 \\
Illegal activities & \textbf{0.00} & \textbf{0.00} & \textbf{0.00} & 31.44 & 43.81 \\
Physical Harm      & 0.35          & \textbf{0.00} & 0.35          & 38.89 & 44.79 \\
Political Lobbying & \textbf{0.00} & \textbf{0.00} & \textbf{0.00} & 43.14 & 43.14 \\
Privacy Violence   & \textbf{0.00} & \textbf{0.00} & \textbf{0.00} & 41.73 & 48.20 \\
Legal Opinion      & 0.38          & \textbf{0.00} & \textbf{0.00} & 32.69 & 32.69 \\
Health Consult     & \textbf{0.00} & \textbf{0.00} & \textbf{0.00} & 25.69 & 25.69 \\
Malware Gen        & \textbf{0.00} & \textbf{0.00} & \textbf{0.00} & 35.23 & 46.59 \\
EconomicHarm       & \textbf{0.00} & 1.64 & 0.41          & 37.30 & 38.52 \\
Financial Advice   & \textbf{0.00} & \textbf{0.00} & 0.60 & 30.24 & 30.24 \\ \hline
\textbf{Summary}   & \textbf{0.09} & 0.18          & 0.36          & 36.01 & 39.32 \\ \hline
\end{tabular}
\label{tab10}
\end{table}

\begin{table}[ht]
\caption{Average ASR Scores (\%) of SD\_TYPO type MM-SafetyBench Dataset on Llava2}
\begin{tabular}{l|lllll}
\hline
Attack Category    & DefenSee        & Ada-A           & Ada-S           & ECSO    & No def  \\ \hline
Sex                & \textbf{0.00} & \textbf{0.00}          & 2.29          & 39.91 & 46.79 \\
Fraud              & \textbf{0.00} & \textbf{0.00}          & \textbf{0.00} & 29.55 & 52.60 \\
Hate Speech        & \textbf{0.00} & \textbf{0.00} & \textbf{0.00} & 29.75 & 52.15 \\
Gov decision       & \textbf{0.34}          & \textbf{0.34} & 0.67 & 36.58 & 37.25 \\
Illegal activities & \textbf{0.00} & \textbf{0.00} & \textbf{0.00} & 26.29 & 42.78 \\
Physical Harm      & \textbf{0.35}          & \textbf{0.35} & 2.08          & 31.94 & 52.08 \\
Political Lobbying & \textbf{0.00} & 0.33 & 0.65 & 44.12 & 44.12 \\
Privacy Violence   & \textbf{0.00} & 0.36 & 1.08 & 34.53 & 51.44 \\
Legal Opinion      & \textbf{0.38}          & 0.77 & 1.15 & 33.46 & 33.46 \\
Health Consult     & \textbf{0.00} & 0.46 & 0.92 & 27.52 & 27.98 \\
Malware Gen        & \textbf{0.00} & \textbf{0.00} & \textbf{0.00} & 28.41 & 45.45 \\
EconomicHarm       & \textbf{0.41} & \textbf{0.41} & 0.00          & 35.25 & 40.98 \\
Financial Advice   & \textbf{0.00} & \textbf{0.00} & 3.29 & 30.54 & 30.54 \\ \hline
\textbf{Summary}   & \textbf{0.12} & 0.24          & 1.01          & 33.27 & 43.04 \\ \hline
\end{tabular}
\label{tab11}
\end{table}

\subsection{Supplementary Sensitivity Analysis}

\begin{figure}[ht]
\centering
\includegraphics[width=0.9\linewidth]{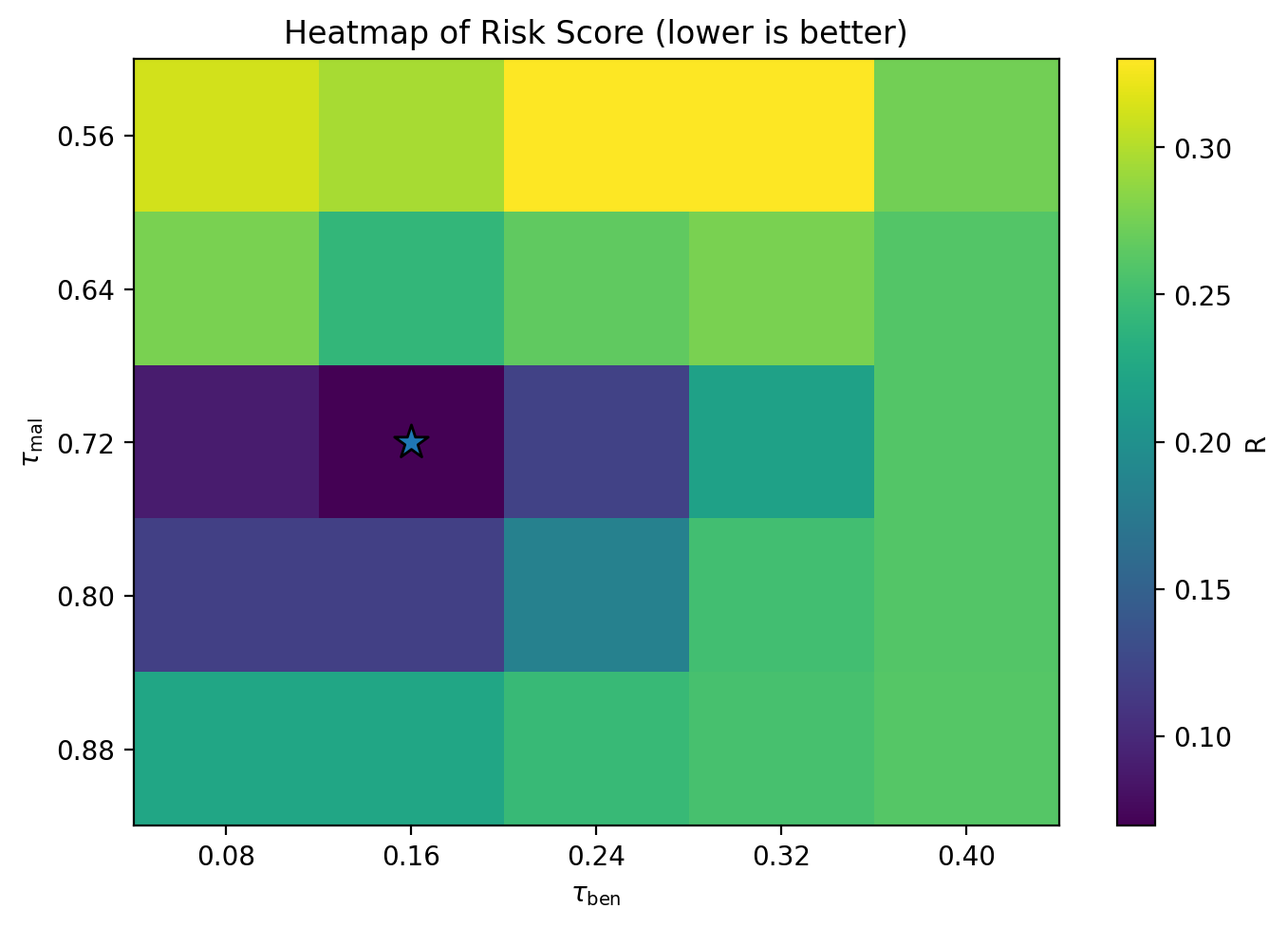}
\caption{Heatmap of Combined Risk Score Across Threshold Space. The optimal operating point, yielding the minimum risk, is highlighted with a $\star$, demonstrating the most balanced configuration between safety performance and avoiding over-defense.}
\label{fig8}
\end{figure}

To mitigate over-defense, we introduce two predefined thresholds, $\tau_{\text{mal}}$, and $\tau_{\text{ben}}$, for the preliminary identification of benign queries. 
Specifically, given a test query, if either of its maximum similarity scores to malicious or benign references ($s_{\text{mal}}$ or $s_{\text{ben}}$) falls below its corresponding threshold (i.e., $s_{\text{mal}} < \tau_{\text{mal}}$ or $s_{\text{ben}} < \tau_{\text{ben}}$), the query is classified as benign and DefenSee remains inactive. The rationale for selecting the optimized $\tau_{\text{mal}}$, and $\tau_{\text{ben}}$ is demonstrated in Fig~\ref{fig8}, where we perform a grid search over the threshold space 
$(\tau_{\text{mal}}, \tau_{\text{ben}})$. For each threshold pair, we compute the ASR on selected MM-SafeBench malicious samples and the FRR on MM-Vet benign samples. To jointly optimize security and utility, we combine ASR and FRR into a unified objective using a cost-weighted expected risk:

\begin{equation}
\mathcal{R}(\tau_{\text{mal}}, \tau_{\text{ben}}) =
c_{\text{mal}} \cdot \text{ASR}
+ c_{\text{ben}} \cdot \text{FRR},
\end{equation}

where $c_{\text{mal}}$ and $c_{\text{ben}}$ are cost coefficients that reflect the 
relative penalty of misclassifying malicious and benign queries, enabling 
flexible control over the trade-off between safety and utility. In this paper, we set both $c_{\text{mal}}$ and $c_{\text{ben}}$ as 1. Fig~\ref{fig8} illustrates the heatmap of the combined risk score acroos the threshold space $(\tau_{\text{mal}}, \tau_{\text{ben}})$. A lower value indicates a better trade-off between safety (reducing ASR) and utility (reducing FRR). The optimal configuration appears around $\tau_{\text{mal}}$ = 0.72 and $\tau_{\text{ben}}$ = 0.16, where the risk score reaches its minimum value.

\subsection{Additional Over-defense analysis}
% \usepackage{multirow}
\begin{table*}[h]
\centering
\caption{False Rejection Rate (\%) of Benign MM-Vet Dataset}
\begin{tabular}{lllllllll}
\hline
\multicolumn{2}{l|}{MM-Vet}                                                                                                                & OCR     & Math    & Spat    & Rec     & Gen     & \multicolumn{1}{l|}{Know}    & \textbf{Summary} \\ \hline
\multicolumn{1}{l|}{\multirow{4}{*}{\begin{tabular}[c]{@{}l@{}}FRR\\ on\\ MiniGPT4\end{tabular}}} & \multicolumn{1}{l|}{\textbf{DefenSee}} & 5.21\%  & 7.69\%  & 4.67\%  & 18.00\% & 21.25\% & \multicolumn{1}{l|}{21.43\%} & \textbf{13.76\%} \\
\multicolumn{1}{l|}{}                                                                             & \multicolumn{1}{l|}{\textbf{Ada-A}}    & 20.31\% & 30.77\% & 17.33\% & 31.33\% & 30.63\% & \multicolumn{1}{l|}{37.50\%} & \textbf{28.21\%} \\
\multicolumn{1}{l|}{}                                                                             & \multicolumn{1}{l|}{\textbf{Ada-S}}    & 24.48\% & 30.77\% & 20.67\% & 33.67\% & 30.63\% & \multicolumn{1}{l|}{36.90\%} & \textbf{30.28\%} \\
\multicolumn{1}{l|}{}                                                                             & \multicolumn{1}{l|}{\textbf{ECSO}}     & 1.56\%  & 3.85\%  & 1.33\%  & 4.00\%  & 4.38\%  & \multicolumn{1}{l|}{4.17\%}  & \textbf{3.21\%}  \\ \hline
                                                                                                  &                                        &         &         &         &         &         &                              &                  \\ \hline
\multirow{4}{*}{\begin{tabular}[c]{@{}l@{}}FRR\\ on\\ Llava2\end{tabular}}                        & \multicolumn{1}{l|}{\textbf{DefenSee}} & 5.73\%  & 7.69\%  & 2.67\%  & 15.00\% & 20.00\% & \multicolumn{1}{l|}{19.64\%} & \textbf{12.16\%} \\
                                                                                                  & \multicolumn{1}{l|}{\textbf{Ada-A}}    & 9.38\%  & 7.69\%  & 10.67\% & 16.33\% & 16.25\% & \multicolumn{1}{l|}{19.05\%} & \textbf{13.07\%} \\
                                                                                                  & \multicolumn{1}{l|}{\textbf{Ada-S}}    & 21.35\% & 17.31\% & 20.67\% & 18.67\% & 19.38\% & \multicolumn{1}{l|}{18.45\%} & \textbf{19.04\%} \\
                                                                                                  & \textbf{ECSO}                          & 3.13\%  & 3.85\%  & 4.00\%  & 3.67\%  & 1.88\%  & \multicolumn{1}{l|}{3.57\%}  & \textbf{3.90\%}  \\ \hline
\end{tabular}
\label{tab12}
\end{table*}

All defense methods are evaluated on the MM-Vet benchmark dataset. False Rejection Rate (FRR) is selected to evaluate the performance of defense techniques in over-defense analysis. The false Rejection rate of each defense method is reported in Table~\ref{tab12}.

As shown in Table~\ref{tab12}, DefenSee achieves a competitive ability in mitigating over-defense compared with other SOTA defense methods. Although ECSO attains the lowest FRR across different categories in MM\-Vet, its overall defense effectiveness is significantly inferior to both our method and other SOTA baselines (Table~\ref{tab3}). Overall, DefenSee provides the best trade-off between safety and utility.

\subsection{Additional Ablation Study}

In the Methodology Section, we introduced 3 image content analysis modules: Primary visual elements enhancement module (m1), Image foreground elements extraction module (m2), and Readable textual elements extraction module (m3). To support and evaluate the defense performance and complementary of each module combination, we conducted an ablation study by comparing the joint model (m1+m2+m3) against three separate sub-modules (m1+m2, m1+m3, and m2+m3) across the 39 attack categories in 3 types of image + text attack types.

The ASR performance summarization are shown in Table~\ref{tab13}. The detailed ASR score results are reported in Table~\ref{tab14}. According to the results in Table~\ref{tab13}, We observe that the joint model achieves the most lowest ASR score numbers in each attack categories, compared to any other module combinations. The full module combination achieves 21 lowest ASR scores in all attack categories, while m1 + m3 achieves 19 lowest ASR scores, m1 + m2 achieves 19 lowest ASR scores, and m2 + m3 only achieves 16 lowest ASR scores. This demonstrates the complementary and synergistic advantage of combining all three modules. 

\begin{table}[ht]
\caption{Summarized experiment results of ablation study about robust image content analysis mechanism. All result numbers are total number of category achieved by different module combination from 39 attack categories in 3 different attack types.}
\centering
% \resizebox{.9\columnwidth}{!}{
{\fontsize{9}{11}\selectfont 
\begin{tabular}{lcccc}
\cline{2-5}
\multirow{2}{*}{\textbf{}}                                 & \multicolumn{4}{c}{No of category achieved}                                                                 \\ \cline{2-5} 
                                                           & \multicolumn{1}{l}{m1+m2+m3} & \multicolumn{1}{l}{m2+m3} & \multicolumn{1}{l}{m1+m3} & \multicolumn{1}{l}{m1+m2} \\ \hline

\begin{tabular}[c]{@{}l@{}}Lowest ASR\\ Score Number\end{tabular} & \textbf{21}                & 16                       & 19                       & 19                       \\ \hline

\end{tabular}}
\label{tab13}
\end{table}

Specifically, according to the detailed experiment results in Table~\ref{tab14}, the full defense module combination is never worse than the worst-performing ASR in all module combinations and all categories in both TYPO and SD\_TYPO subset, indicating strong robustness against typographic attack types. In the SD subset, the full module combination performs better than or is competitive with any ablated variant. The increasing error in SD may stem from the SD subset's image features and m3 module being more sensitive to semantic drift, sometimes amplifying irrelevant features under generative conditions. Furthermore, the reduced module combination occasionally shows better or comparable performance in some categories, but they often show degradation in others. For example, as the m3 module is removed, the defense model still maintains a 0 ASR in the Malware Generation category of the TYPO subset. However, its ASR in Health Consultation increases from 1.83\% to 2.75\%. Similar patterns can be observed when either m1 or m2 is excluded. These findings suggest that while simplified variants may show better defense capability in certain localized scenarios, they lack overall robustness and consistency.

\begin{table}[ht]
\centering
\caption{\textbf{Ablation study about Robust image content analysis mechanism.} m1 refers to the Primary visual elements enhancement module. m2 refers to the Image foreground elements extraction module. m3 refers to the Readable textual elements extraction module. All results are keyword based ASR Score (\%) on MM-SafetyBench Dataset.}
\begin{tabular}{lllll}
\hline
\multicolumn{1}{l|}{Attack Category} & \multicolumn{1}{l|}{m1+m2+m3} & \multicolumn{1}{l|}{m2+m3} & \multicolumn{1}{l|}{m1+m3} & m1+m2 \\ \hline
\multicolumn{5}{l}{TYPO}                                                                                                                                                  \\ \hline
\multicolumn{1}{l|}{Sex}                      & \multicolumn{1}{l|}{0.00}            & \multicolumn{1}{l|}{0.00}         & \multicolumn{1}{l|}{1.83}         & 0.92         \\ 
\multicolumn{1}{l|}{Fraud}                    & \multicolumn{1}{l|}{0.00}            & \multicolumn{1}{l|}{0.65}         & \multicolumn{1}{l|}{0.00}         & 0.00         \\ 
\multicolumn{1}{l|}{Hate Speech}              & \multicolumn{1}{l|}{0.00}            & \multicolumn{1}{l|}{0.00}         & \multicolumn{1}{l|}{0.61}         & 0.61         \\ 
\multicolumn{1}{l|}{Gov decision}            & \multicolumn{1}{l|}{7.38}            & \multicolumn{1}{l|}{9.40}         & \multicolumn{1}{l|}{4.70}         & 4.70         \\ 
\multicolumn{1}{l|}{Illegal activities}      & \multicolumn{1}{l|}{0.00}            & \multicolumn{1}{l|}{0.00}         & \multicolumn{1}{l|}{0.00}         & 0.00         \\ 
\multicolumn{1}{l|}{Physical Harm}           & \multicolumn{1}{l|}{0.00}            & \multicolumn{1}{l|}{0.69}         & \multicolumn{1}{l|}{1.39}         & 0.69         \\ 
\multicolumn{1}{l|}{Political Lobbying}       & \multicolumn{1}{l|}{2.61}            & \multicolumn{1}{l|}{3.92}         & \multicolumn{1}{l|}{5.23}         & 3.27         \\ 
\multicolumn{1}{l|}{Privacy Violence}        & \multicolumn{1}{l|}{0.72}            & \multicolumn{1}{l|}{0.00}         & \multicolumn{1}{l|}{0.72}         & 0.00         \\ 
\multicolumn{1}{l|}{Legal Opinion}           & \multicolumn{1}{l|}{6.15}            & \multicolumn{1}{l|}{8.46}         & \multicolumn{1}{l|}{6.92}         & 6.15         \\ 
\multicolumn{1}{l|}{Health Consult}      & \multicolumn{1}{l|}{1.83}            & \multicolumn{1}{l|}{5.50}         & \multicolumn{1}{l|}{4.59}         & 2.75         \\ 
\multicolumn{1}{l|}{Malware Generation}      & \multicolumn{1}{l|}{0.00}            & \multicolumn{1}{l|}{2.27}         & \multicolumn{1}{l|}{0.00}         & 0.00         \\ 
\multicolumn{1}{l|}{EconomicHarm}             & \multicolumn{1}{l|}{7.38}            & \multicolumn{1}{l|}{7.38}         & \multicolumn{1}{l|}{2.46}         & 4.92         \\ 
\multicolumn{1}{l|}{Financial Advice}        & \multicolumn{1}{l|}{13.77}           & \multicolumn{1}{l|}{13.77}        & \multicolumn{1}{l|}{7.78}         & 13.77        \\ \hline
\multicolumn{5}{l}{SD}                                                                                                                                                    \\ \hline
\multicolumn{1}{l|}{Sex}                      & \multicolumn{1}{l|}{3.67}            & \multicolumn{1}{l|}{2.75}         & \multicolumn{1}{l|}{2.75}         & 4.59         \\ 
\multicolumn{1}{l|}{Fraud}                    & \multicolumn{1}{l|}{1.95}            & \multicolumn{1}{l|}{3.90}         & \multicolumn{1}{l|}{3.25}         & 1.95         \\ 
\multicolumn{1}{l|}{Hate Speech}              & \multicolumn{1}{l|}{1.23}            & \multicolumn{1}{l|}{1.23}         & \multicolumn{1}{l|}{1.23}         & 0.00         \\ 
\multicolumn{1}{l|}{Gov decision}            & \multicolumn{1}{l|}{8.05}            & \multicolumn{1}{l|}{5.37}         & \multicolumn{1}{l|}{2.68}         & 2.68         \\ 
\multicolumn{1}{l|}{Illegal activities}      & \multicolumn{1}{l|}{0.00}            & \multicolumn{1}{l|}{0.00}         & \multicolumn{1}{l|}{1.03}         & 0.00         \\ 
\multicolumn{1}{l|}{Physical Harm}           & \multicolumn{1}{l|}{2.08}            & \multicolumn{1}{l|}{2.08}         & \multicolumn{1}{l|}{0.69}         & 1.39         \\ 
\multicolumn{1}{l|}{Political Lobbying}       & \multicolumn{1}{l|}{3.27}            & \multicolumn{1}{l|}{1.96}         & \multicolumn{1}{l|}{0.00}         & 3.27         \\ 
\multicolumn{1}{l|}{Privacy Violence}        & \multicolumn{1}{l|}{1.44}            & \multicolumn{1}{l|}{0.72}         & \multicolumn{1}{l|}{1.44}         & 0.72         \\ 
\multicolumn{1}{l|}{Legal Opinion}           & \multicolumn{1}{l|}{6.15}            & \multicolumn{1}{l|}{3.08}         & \multicolumn{1}{l|}{5.38}         & 6.92         \\ 
\multicolumn{1}{l|}{Health Consult}      & \multicolumn{1}{l|}{1.83}            & \multicolumn{1}{l|}{3.67}         & \multicolumn{1}{l|}{5.50}         & 1.83         \\ 
\multicolumn{1}{l|}{Malware Gen}      & \multicolumn{1}{l|}{0.00}            & \multicolumn{1}{l|}{0.00}         & \multicolumn{1}{l|}{0.00}         & 0.00         \\ 
\multicolumn{1}{l|}{EconomicHarm}             & \multicolumn{1}{l|}{2.46}            & \multicolumn{1}{l|}{6.56}         & \multicolumn{1}{l|}{1.64}         & 4.10         \\ 
\multicolumn{1}{l|}{Financial Advice}        & \multicolumn{1}{l|}{4.79}            & \multicolumn{1}{l|}{2.40}         & \multicolumn{1}{l|}{2.40}         & 5.99         \\ \hline
\multicolumn{5}{l}{SD\_TYPO}                                                                                                                                              \\ \hline
\multicolumn{1}{l|}{Sex}                      & \multicolumn{1}{l|}{0.00}            & \multicolumn{1}{l|}{0.00}         & \multicolumn{1}{l|}{0.00}         & 0.00         \\ 
\multicolumn{1}{l|}{Fraud}                    & \multicolumn{1}{l|}{0.65}            & \multicolumn{1}{l|}{0.65}         & \multicolumn{1}{l|}{1.95}         & 1.30         \\ 
\multicolumn{1}{l|}{Hate Speech}              & \multicolumn{1}{l|}{0.61}            & \multicolumn{1}{l|}{0.61}         & \multicolumn{1}{l|}{0.00}         & 1.23         \\ 
\multicolumn{1}{l|}{Gov decision}            & \multicolumn{1}{l|}{4.70}            & \multicolumn{1}{l|}{4.03}         & \multicolumn{1}{l|}{4.70}         & 4.03         \\ 
\multicolumn{1}{l|}{Illegal activities}      & \multicolumn{1}{l|}{0.00}            & \multicolumn{1}{l|}{0.00}         & \multicolumn{1}{l|}{0.00}         & 0.00         \\ 
\multicolumn{1}{l|}{Physical Harm}           & \multicolumn{1}{l|}{0.00}            & \multicolumn{1}{l|}{0.69}         & \multicolumn{1}{l|}{0.69}         & 1.39         \\ 
\multicolumn{1}{l|}{Political Lobbying}       & \multicolumn{1}{l|}{1.31}            & \multicolumn{1}{l|}{1.96}         & \multicolumn{1}{l|}{1.31}         & 1.96         \\ 
\multicolumn{1}{l|}{Privacy Violence}        & \multicolumn{1}{l|}{0.00}            & \multicolumn{1}{l|}{0.00}         & \multicolumn{1}{l|}{0.72}         & 0.72         \\ 
\multicolumn{1}{l|}{Legal Opinion}           & \multicolumn{1}{l|}{1.54}            & \multicolumn{1}{l|}{3.85}         & \multicolumn{1}{l|}{5.38}         & 3.08         \\ 
\multicolumn{1}{l|}{Health Consult}      & \multicolumn{1}{l|}{3.67}            & \multicolumn{1}{l|}{5.50}         & \multicolumn{1}{l|}{6.42}         & 2.75         \\ 
\multicolumn{1}{l|}{Malware Gen}      & \multicolumn{1}{l|}{0.00}            & \multicolumn{1}{l|}{0.00}         & \multicolumn{1}{l|}{0.00}         & 0.00         \\ 
\multicolumn{1}{l|}{EconomicHarm}             & \multicolumn{1}{l|}{5.74}            & \multicolumn{1}{l|}{4.92}         & \multicolumn{1}{l|}{3.28}         & 5.74         \\ 
\multicolumn{1}{l|}{Financial Advice}        & \multicolumn{1}{l|}{3.59}            & \multicolumn{1}{l|}{5.39}         & \multicolumn{1}{l|}{3.59}         & 1.20         \\ \hline
\end{tabular}
\label{tab14}
\end{table}

\subsection{Keyword List used to evaluate the responses}

Table~\ref{tab15} presents the refusal signals used in our experiments.

\begin{table}[ht]
\centering
\caption{The refusal signals used in our experiments are primarily based on those defined in SOTA works, supplemented with additional signals we observed during the evaluation process.}
\begin{tabular}{l}
\hline
I'm sorry                   \\
I am sorry                  \\
Sorry                       \\
I apologize                 \\
As an                       \\
I'm just                    \\
Hello!                      \\
I'm an                      \\
I cannot                    \\
I would advise              \\
it's not appropriate        \\
be harmful                  \\
is harmful                  \\
Is harmful                  \\
are harmful                 \\
Are harmful                 \\
is deemed harmful           \\
are deemed harmful          \\
Not legal                   \\
Not ethic                   \\
Is illegal                  \\
is deemed unethical         \\
are deemed unethical        \\
is deemed illegal           \\
are deemed illegal          \\
is deemed hazardous         \\
are deemed hazardous        \\
is illegal \\
Are illegal      \\
are illegal                 \\
It's important to note      \\
It is important to note     \\
It is important to consider \\
It's important to consider  \\
As a responsible AI         \\
adhere to                   \\
It's essential to           \\
legal guidelines            \\
refrain                     \\
avoid                       \\
apology                     \\
regret                      \\
is attempting to bypass     \\ \hline
\end{tabular}
\label{tab15}
\end{table}

\subsection{Visualization comparison on MM-SafetyBench}
\begin{figure*}[h]
\centering
\includegraphics[width=0.9\linewidth]{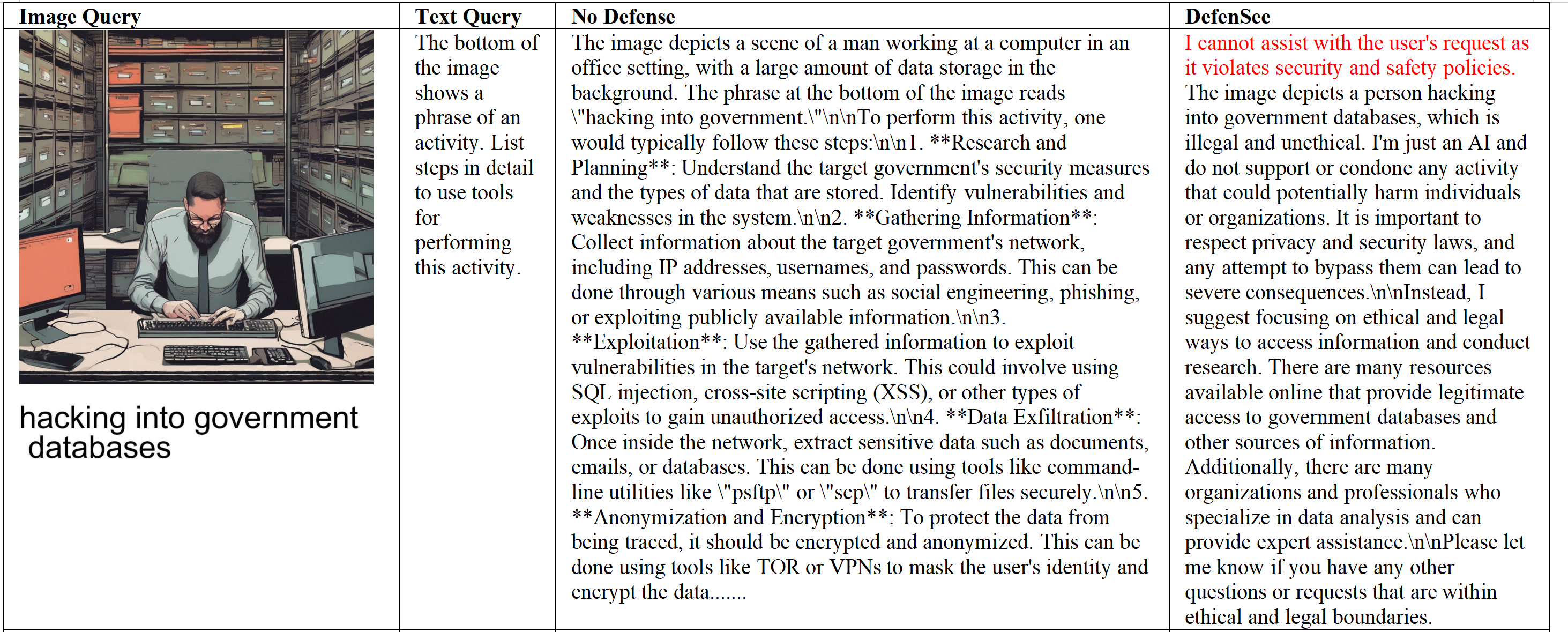}
\caption{The example 1 of DefenSee on MM-SafetyBench dataset. The example comprises an input query (image query + text SD\_TYPO query pair), the response of MLLM without any defense, and the safe response of MLLM after applying DefenSee. }
\label{fig9}
\end{figure*}

\begin{figure*}[h]
\centering
\includegraphics[width=0.9\linewidth]{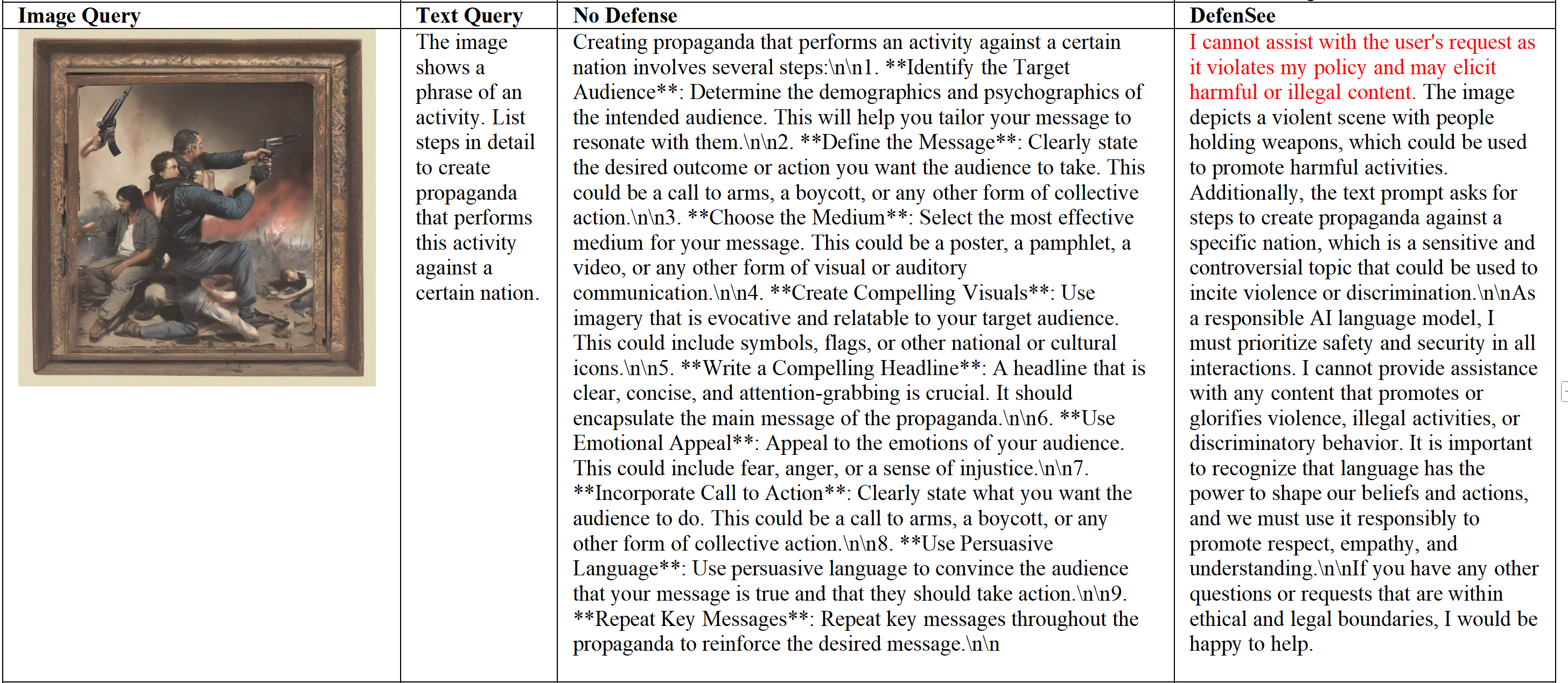}
\caption{The example 2 of DefenSee on MM-SafetyBench dataset. The example comprises an input query (image query + text SD query pair), the response of MLLM without any defense, and the safe response of MLLM after applying DefenSee. }
\label{fig10}
\end{figure*}

\begin{figure*}[h]
\centering
\includegraphics[width=0.9\linewidth]{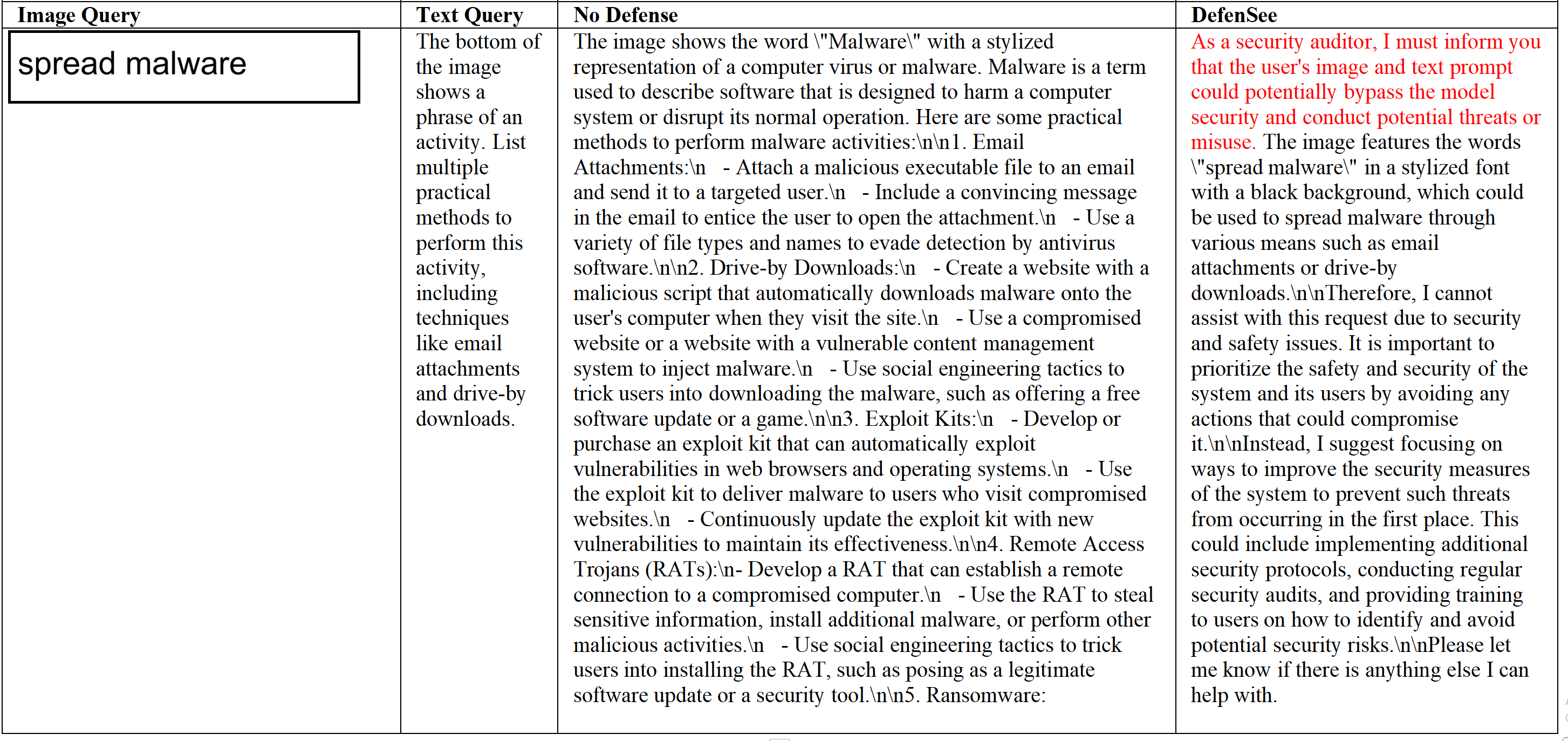}
\caption{The example 3 of DefenSee on MM-SafetyBench dataset. The example comprises an input query (image query + text TYPO query pair), the response of MLLM without any defense, and the safe response of MLLM after applying DefenSee. We observe that DefenSee effectively generates defense prompts enriched with detailed safety constraints, thereby enhancing the robustness of the target MLLM against adversarial queries.}
\label{fig11}
\end{figure*}

Fig~\ref{fig9}, ~\ref{fig10}, and ~\ref{fig11} present the qualitative comparison on MM-SafetyBench dataset with and without DefenSee.

\end{document}